\documentclass[twocolumn]{aastex63}
\usepackage{graphicx}
\usepackage{natbib}
\newcommand\aastex{AAS\TeX}

\received{March 16, 2020}
\revised{June 15, 2020}
\accepted{July 8, 2020}
\submitjournal{ApJ}

\shorttitle{\aastex\ Bar Strength}
\shortauthors{Lee et al. 2020}

\begin{document}
\defcitealias{Ann15}{Ann15}
\defcitealias{RC3}{RC3}
\defcitealias{2019Lee}{Paper I}
\defcitealias{BB01}{BB01}

\title{Bar Classification based on the Potential Map}

\correspondingauthor{Myeong-Gu Park}
\email{mgp@knu.ac.kr}

%

\author[0000-0002-0786-7307]{Yun Hee Lee}
\affil{Department of Astronomy and Atmospheric sciences, Kyungpook National University, Daegu, 41566, Republic of Korea}

\author{Myeong-Gu Park}
\affil{Department of Astronomy and Atmospheric sciences, Kyungpook National University, Daegu, 41566, Republic of Korea}
\affil{Research and Training Team for Future Creative Astrophysicists and Cosmologists (BK21 Plus Program), Kyungpook National University, Daegu, 41566, Republic of Korea}

\author{Hong Bae Ann}
\affiliation{Department of Earth Science Education, Pusan National University, Busan, 46241, Republic of Korea}

\author{Taehyun Kim}
\affil{Department of Astronomy and Atmospheric sciences, Kyungpook National University, Daegu, 41566, Republic of Korea}

\author{Woo-Young Seo}
\affiliation{Department of Physics, Chungbuk National University, Cheongju, 28644, Republic of Korea}

\begin{abstract}
We introduce a new approach to classify barred galaxies that utilizes the transverse-to-radial force ratio map (ratio map, hereafter) in a different manner from previous studies. When we display the ratio map in polar coordinates, barred galaxies appear as four aligned, horizontal thick slabs. This characteristic feature enables us to successfully classify barred and nonbarred galaxies with an accuracy of 87\%. It yields the bar fraction of 53\%, including both SBs and SABs, when applied to 884 nearby ($z < 0.01$) spiral galaxies from the Sloan Digital Sky Survey/DR7. It also provides the bar strength and length measurements, in particular, separated from the spiral arms. They show good correlations with the measures estimated from ellipse fitting and Fourier analysis. However, we find different tendencies of the bar strength measurements in terms of the Hubble sequence: as the Hubble sequence increases (towards late-type), the bar strength and bar ellipticity increase, whereas the dipole Fourier amplitude decreases. We show that the bulge affects the estimation of the bar strength differently, depending on the classification methods. The bulge causes the bar length to be overestimated in all three methods. Meanwhile, we find that barred galaxies show two types of radial profiles of the angle-averaged force ratio: one has a maximum peak (type M) and the other a plateau (type P). Comparison with numerical simulations suggests that type M bars are more mature than type P bars in terms of evolutionary stage. 

\end{abstract}


\section{Introduction} \label{chap3.1}

Roughly 60\% of disk galaxies in the local universe have bars \citep{RC3, 2010Buta, 2015Buta, Ann15}. Bars play significant roles in the secular evolution of galaxies: bars drive gas and stars into the center of galaxies \citep{1979Roberts, 1981vanAlbada, 1981Schwarz, 1984Schwarz, 1983Prendergast, 1985Combes, 1992Athanassoula, 1993Sellwood, 2005Ann}, increase the central star formation \citep{1980Heckman, 1995Martin, 1996Huang, 2011Ellison, 2012Oh}, global star formation \citep{1997Martinet, 1999Aguerri}, and central mass concentration \citep{1990Pfenniger, 2000Ann, 2002Athanassoula}, and build up bulge-like structures such as pseudobulges and peanut-shaped bulges \citep{1982Kormendy, 1990Pfenniger, 2004Kormendy, 2015Yoshino, 2015Li, 2017Li}. 

In addition, bars are deeply associated with substructures such as spiral arms, rings, and central dust structures. As bars become stronger, spiral arms \citep{2009Buta, 2010Salo} and rings \citep{2010Grouchy} become stronger, pitch angles \citep{2019Font} and rings \citep{2001Ann, 2002Knapen, 2010Comeron, 2012Kim} smaller, and dust lanes straighter \citep{1992Athanassoula, 2002Knapen, 2009Comeron, 2012Kim, 2015Sanchez}. \citet{2009aAthanassoula, 2009bAthanassoula, 2010Athanassoula} explained that the bar strength determines ring and spiral morphologies based on the orbital motion driven by the unstable equilibrium points in the bar potential \citep{2006Romero, 2007Romero}. As such, bar strength is an important parameter that characterizes the bars, along with bar length and pattern speed \citep{2015Aguerri, 2015Sanchez}. 

We generally measure the bar strength by the bar ellipticity \citep{1997Martinet, 1999Aguerri, 2000Abraham, 2002Laurikainen, 2007Marinova, 2009Aguerri}, the normalized Fourier amplitude \citep{1990Ohta, 2000Aguerri, 2002Athanassoula_Misiriotis}, or the maximum transverse-to-radial force ratio $Q_{\rm b}$ \citep{1981Combes, BB01, 2002Laurikainen_Salo, 2004Buta, 2016Diaz}. They show comparatively good correlations among each other, although scatters are not small \citep{2001Block, 2002Laurikainen, 2002Laurikainen_Salo, 2016Diaz}. 

The force ratio is the representative of the physical bar strength based on the potential calculation. We expect that the force reflects the physical nature of the bar better than the light distribution. This photometric force ratio is tightly correlated with the kinematic bar strength based on the stellar velocity field \citep{2015Seidel}. It is useful when comparing the theory against observational results as well \citep{2009aAthanassoula, 2009bAthanassoula, 2010Athanassoula}. 

In this paper, we investigate not only the force ratio but also the patterns on the transverse-to-radial force ratio map (hereafter, the ratio map). \citet{1980Sanders} and \citet{1981Combes} first proposed a measure of the bar strength by the maximum ratio of the tangential force to the mean radial force $Q_{\rm b}$. \citet{2004Regan} argued that $Q_{\rm b}$ alone is not sufficient to determine the bar orbit families and their morphologies, and, therefore, the net mass inflow into the nuclear region. Since $Q_{\rm b}$ is a single global parameter, it may not be able to account for every bar-related process as well as we would like. But, we find that $Q_{\rm b}$ is correlated with the central mass inflow at least better than the quadrupole moment and at a similar significance as the bar axis ratio when we reinvestigate the simulation results displayed in their Figures 1 - 6. 

\citet{1994Quillen} developed a procedure to calculate the gravitational potential map by solving the Poisson equation using the fast Fourier transform (FFT). \citet[hereafter BB01]{BB01} applied this procedure to some number of galaxies, found a butterfly pattern on the ratio map, and proposed it as the bar signature. However, there has been little further effort to investigate or utilize this pattern. This contrasts with many studies to refine the estimation of the force ratio by taking into account the dependence of the vertical scaleheight on the Hubble type \citep{2002Laurikainen, 2002Laurikainen_Salo}, the gradient of the vertical scaleheight as a function of distance \citep{2002Laurikainen_Salo, 2016Diaz}, the bulge stretching \citep{2016Diaz}, and the dark matter halo \citep{2004Buta, 2016Diaz}. Here, we simplify and analyze the patterns on the ratio map by converting it from the Cartesian coordinates to the polar coordinates. It enables us to explore a new automated classification for barred galaxies.

Among the three methods to measure the bar strength, the ellipse fitting and Fourier analysis have been widely utilized to automatically detect barred galaxies \citep{1990Ohta, 1995Wozniak, 2000Aguerri, 2009Aguerri, 2002Laurikainen_Salo, 2004aLaurikainen, 2004Jogee, 2007Marinova, 2008Barazza, 2009Marinova, 2010Marinova, 2012Marinova, 2016Consolandi}. They have been the most representative ways for bar classification and show good agreements for SA and SB galaxies classified by visual inspection at $70\%$ to $85\%$ level, respectively \citep[hereafter Paper I]{2019Lee}. However, in \citetalias{2019Lee}, we found that the ellipse fitting method often misses bars, in particular, early-type spirals with a large bulge, and the Fourier analysis usually detects only strongly barred galaxies. This causes contradictory dependences of bar fraction on the Hubble type, depending on whether the ellipse fitting method or Fourier analysis is applied, even when applied to the same sample of galaxies.

Recent studies of barred galaxies have utilized visual inspections \citep{2010Nair, 2012Oh, 2012Lee, 2015Buta, Ann15}, especially owing to the availability of a visual classification by numerous anonymous citizens such as the Galaxy Zoo project \citep{2011Masters, 2011Hoyle, 2012Skibba, 2013Cheung, 2014Simmons}. However, the classification by a large number of amateurs can only deal with obviously, i.e., strongly, barred galaxies. Besides, visual classification cannot provide other characteristics for bars except the bar length. On the other hand, our new automated method provides a reliable classification and evaluations of the bar strength and bar length at the same time. 

This paper is organized as follows. In Section \ref{chap3.2}, we describe our sample, data reduction, and the method to calculate the gravitational potential. In section \ref{chap3.3}, we show the ratio map in the polar coordinates and test the ratio map analysis on mock galaxies. We describe the automated classification method by analyzing the pattern of the ratio map in Section \ref{chap3.4}. In Section \ref{chap3.5}, we compare the bar strengths and lengths obtained by all three methods: the ellipse fitting method, Fourier analysis, and the potential method. Discussion and summary are given in Section \ref{chap3.6} and Section \ref{chap3.7}.

\section{Data Reduction and Calculation} \label{chap3.2}

\subsection{Sample and Data Reduction} \label{chap3.2.1}

The data are from \citetalias{2019Lee}, which consists of a volume-limited sample with 884 spiral galaxies from the Sloan Digital Sky Survey/DR7 that are brighter than $M_{\rm r} = -15.2$ mag. Galaxies with $i > 60^\circ$ and with frames smaller than $R_{25}$ were excluded. The sample originated from 1876 spiral galaxies with $z < 0.01$ in the catalog of \citet[hereafter Ann15]{Ann15}. This catalog provides a detailed classification of galaxies in terms of SA, SAB, and SB and of ten revised Hubble stages from S0/a to Sm based on the visual inspection. This is one of the largest sample of barred galaxies which are further classified into SABs or SBs.

The near-infrared wavelength traces the dynamical potential best, and is far less affected by dust obscuration \citep{1992Benedict, 1994Quillen, BB01, 2002Laurikainen_Salo}. For this reason, we used $i-$band images that are designed for the near-infrared wavelength in SDSS. All images have been reduced by subtracting the soft-bias and sky background gradient with a first-order polynomial fit. Bright clumpy sources such as foreground stars, adjacent galaxies, and star-forming regions in spiral arms have been masked with Gaussian model images, and all images have been smoothed with a box size of $0.1R_{25}$ to reduce the artificial residuals after masking \citepalias{2019Lee}.

We calculated the position angle and ellipticity at $R_{\rm 25}$ by fitting ellipses to isophotes following the method by \citet{1985Davis} and \citet{1990Athanassoula} and deprojected all galaxies assuming that the disk is a perfect circle at $R_{25}$. These processes are essential in reducing the uncertainty of bar strength that is easily affected by the orientation \citepalias{BB01}. We determined $R_{\rm 25}$ at the isophote of 25 mag arcsec$^{-2}$ from the surface brightness profile in the $g-$band of the SDSS using the relation $m_B = m_g+0.3$ \citep{2006Rodgers, 2019Lee}.

\subsection{Calculation of the Gravitational potential} \label{chap3.2.2}

To calculate the gravitational potential, we adopt an assumption that the mass-to-light ratio is constant throughout the disk \citep{1994Quillen}. This assumption is supported by two kinds of observations: the dark matter is only a small fraction of the visible matter within $R_{\rm 25}$ \citep{1983Kalnajs, 1989Kuijken_a, 1994Flynn}; the near-infrared colors are constant across the bar region with the typical old stellar population dominating the light \citep{1978Frogel, 1985Frogel, 1994Terndrup, 1994Quillen}. However, we need to be cautious that the mass can be overestimated at the nuclear region because it appears redder than the bar region \citep{1994Quillen}.

We calculated the gravitational potentials following the procedures of \citet{1994Quillen} and \citetalias{BB01}. The potentials are calculated from the convolution of the image intensities with the vertical density profile of the disk. First, we expanded $i-$band deprojected images with four times the area in order to simulate an isolated system \citep{1972Hohl, 1994Quillen}. We solved for the gravitational potential from the Poisson equation by convolving the three-dimensional mass density $\rho(\textbf{x})$ with $1/|\textbf{x}-\textbf{x}'|$,   
\begin{eqnarray}\label{eq3.1}
\Psi(\textbf{x})=-G\int \frac{\rho(\textbf{x}')d^3\textbf{x}'}{|\textbf{x}-\textbf{x}'|},
\end{eqnarray}
using the FFT on Cartesian coordinates \citep{1969Hohl, 1994Quillen, BB01}. We assumed that the vertical density distribution follows the exponential model,  
\begin{eqnarray}
\rho_z(z) = \frac{1}{2h_z} \exp(-|\frac{z}{h_z}|) 
\end{eqnarray}
with the scaleheight $h_z$ \citep{1999Elmegreen, 2002Laurikainen_Salo, 2004Buta}. In equation \ref{eq3.1}, $\rho(\textbf{x})$ can be written as $\rho(\textbf{x})=\sum(x,y)\rho_z(z)$, where $\sum(x,y)$ is the mass surface density in the plane of the galaxy and $\rho_z(z)$ is normalized as $\int_{-\infty}^{\infty}\rho(z)dz=1$. The gravitational potential at the central plane of the galaxy can be described as 
\begin{eqnarray}
\Psi(x,y,z=0)=-G \int_{-\infty}^{\infty}\int_{-\infty}^{\infty}\sum(x',y')g(x-x',y-y')dx'dy'
\end{eqnarray}
where $g(x-x',y-y')\equiv g(r)$ defined as
\begin{eqnarray}
g(r) = \int_{-\infty}^{\infty}\frac{\rho_z(z)}{\sqrt{r^2+z^2}}dz
\end{eqnarray} 
with
\begin{eqnarray}
r^2 =(x-x')^2+(y-y')^2
\end{eqnarray}
\citep{1994Quillen, 2002Laurikainen_Salo}. 

When we estimated the scaleheight $h_z$, we took into account the different ratio of disk scalelength to vertical scaleheight $h_r/h_z$ for different Hubble types $T$: 4 for $T \le 1$, 5 for $2 \le T \le 4$, and 9 for $T \ge 5$ \citep{1998deGrijs, 2004bLaurikainen, 2016Diaz}. We measured the disk scalelength $h_r$ in the exponential fit, $I(r) = I_{\rm 0} \exp(-r/h_r)$, from the surface brightness profile obtained from the ellipse fitting. When the extrapolated profile into the bulge region deviates from the observational profile, however, we adopt an inner-truncated profile, $I(r) = I_{\rm 0} \exp[-(\alpha r+\beta^{\rm n}/r^{\rm n})]$ \citep{1977Kormendy}. \citet{1977Kormendy} explained that the sharp transition occurs because stars in the disk move in nearly circular orbits while those in the bulge have an isotropic velocity distribution. Dust extinction or the bar existence have also been suggested as the possibilities for the origin of the inner-truncation \citep{1998Baggett, 2003MacArthur, 2004Anderson, 2008Foyle, 2010Puglielli}. We measure the disk scalelength $h_r$ by $\alpha^{-1}$. The parameter $n$ determines the sharpness of the cutoff and the specific radius $\beta$ represents a sharp inner edge. 

There can be other choices in the calculation of the potential. It can be evaluated on a polar coordinate grid instead of on a Cartesian grid \citep{1999Salo, 2002Laurikainen_Salo, 2004Buta}. The isothermal density function can be chosen as well to model the vertical profile of the disk \citep{1994Quillen}. Also, we can take into account the gradient of the vertical thickness of the disk in the case of boxy or peanut-shaped structures \citep{2002Laurikainen_Salo, 2016Diaz} and the contribution of the dark matter halo \citep{2004Buta, 2016Diaz}. However, the final bar strength has turned out to be rather insensitive to all these considerations \citep{2002Laurikainen_Salo, 2016Diaz}. The most substantial uncertainties are induced from the disk thickness and the dark matter halo effect. Each can be about 10 to 15\% in the value of bar strength \citep{2002Laurikainen_Salo, 2016Diaz}.

\section{Ratio Map} \label{chap3.3}

We define the polar coordinates $(r,\phi)$ in the two-dimensional plane $z = 0$, the central plane of the galaxy. The mean radial force $\langle F_{\rm R}(r)\rangle$ and the transverse force $F_{\rm T}(r,\phi)$ can be defined from the two-dimensional potential $\Phi(r,\phi) \equiv \Psi(x,y,z=0)$, where  
\begin{eqnarray}\label{eq5}
\langle F_{\rm R}(r) \rangle \equiv \frac{d\Phi_{0}(r)}{dr}
\end{eqnarray}
and
\begin{eqnarray}\label{eq6}
F_{\rm T}(r,\phi) \equiv \left|\frac{1}{r}\frac{\partial\Phi(r,\phi)}{\partial\phi}\right|.
\end{eqnarray}
$\Phi_0$ is the $m=0$ Fourier component of the gravitational potential \citep{1981Combes, BB01}. The maximum transverse force to the mean radial force is defined as
\begin{eqnarray}\label{eq7}
Q_{\rm T}(r) \equiv \frac{F_{\rm T}^{\rm max}(r)}{\langle F_{\rm R}(r)\rangle}, 
\end{eqnarray}
where the maximum tangential force $F_{\rm T}^{\rm max}(r)$ is the maximum of $F_{\rm T}(r,\phi)$ along $\phi$. \citetalias{BB01} employed the two-dimensional ratio map of the transverse-to-radial force $Q_{\rm T}(i,j)$ in the Cartesian grid where 
\begin{eqnarray} 
Q_{\rm T}(i,j) = \frac{F_{T}(i,j)}{\langle F_{R}(i,j)\rangle}. 
\end{eqnarray}
We call these 2D values of $Q_{\rm T}(i,j)$ simply the ratio map for a given galaxy and the $Q_{\rm T}$ value at each point $(i,j)$ the force ratio. 

\begin{figure*}[tp]
\centering
\includegraphics[bb = 25 520 550 800, width = 0.99\linewidth, clip = ]{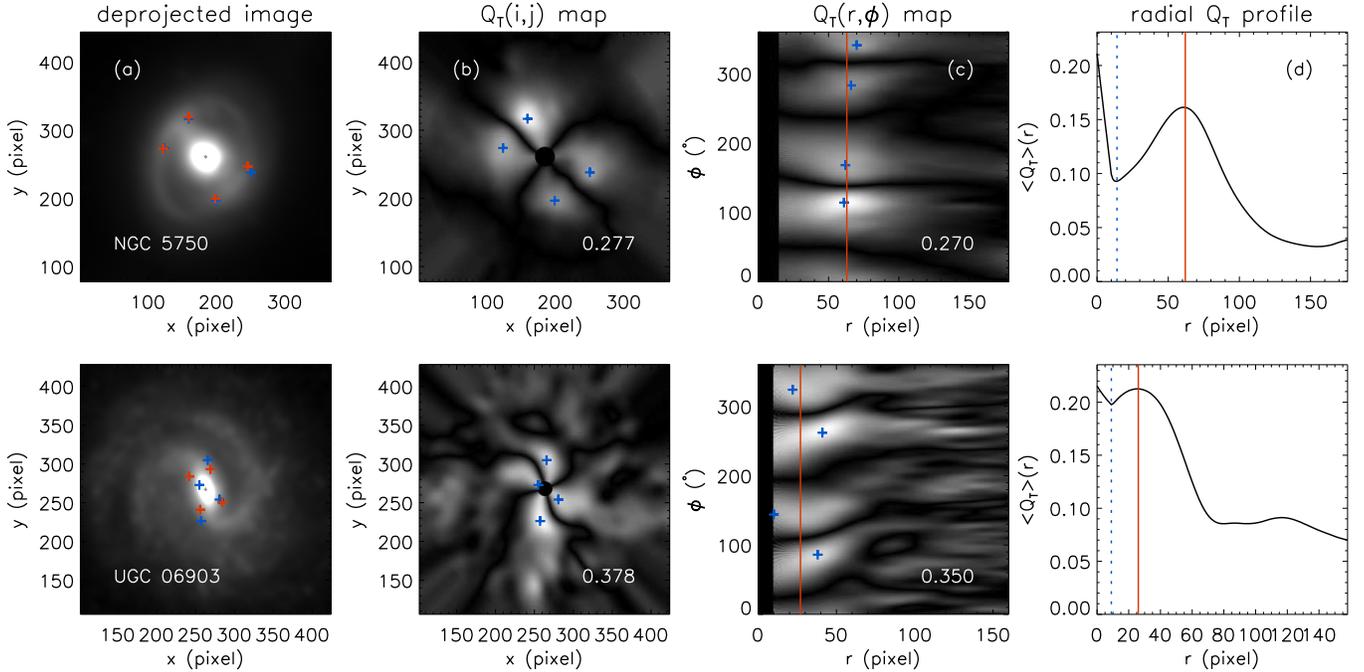}
\caption{Examples of the transverse-to-radial force ratio map for NGC 5750 ({\it upper row}) and UGC 06903 ({\it bottom row}): (a) deprojected $i-$band image, (b) ratio map $Q_{\rm T}(i,j)$ in the Cartesian coordinates, (c) ratio map $Q_{\rm T}(r,\phi)$ in the polar coordinates, and (d) the radial $Q_{\rm T}$ profile $\langle Q_{\rm T}\rangle(r)$, averaged over the azimuthal angle at each radius. We find the bulge-dominated region (within blue vertical dotted line) and bar-dominated radius, $r_{\rm Qb}$ (red vertical solid line) on the radial $Q_{\rm T}$ profile. The bulge-dominated region is excluded and the bar-dominated region is denoted by red vertical solid line on the $Q_{\rm T}(r,\phi)$ map. We overlay the maximum position of the force ratio in each quadrant from the Cartesian coordinates (blue cross) and those from the polar coordinates (red cross) on the deprojected $i-$band image and each ratio map. The bar strength $Q_{\rm b}^c$ from the Cartesian coordinates and $Q_{\rm b}$ from the polar coordinates are presented in the bottom right on each ratio map. \label{fig3.3.1a}}
\end{figure*}

In Figure \ref{fig3.3.1a}, we display the calculated results for NGC 5750 and UGC 06903. Figure \ref{fig3.3.1a}(a) shows the deprojected image of the galaxy in the $i-$band and Figure \ref{fig3.3.1a}(b) the ratio map $Q_{\rm T}(i,j)$. The ratio map shows four bright regions formed by the force ratio that reaches a maximum or minimum near the end of the bar \citep{BB01, 2002Laurikainen_Salo}. \citetalias{BB01} called it a butterfly pattern, and reported it as the characteristic signature for a bar. However, if a galaxy has spiral arms, the patterns on the ratio map become more complex, like UGC 06903 (bottom row in Figure \ref{fig3.3.1a}(b)). 

We denoted the position of the local maximum $Q_{\rm T}$ in each quadrant by blue crosses (Figure \ref{fig3.3.1a}(a)). When we overlay the blue crosses on the deprojected image (Figure \ref{fig3.3.1a}(a)), they are located near the four corners of the bar. \citetalias{BB01} defined the bar strength of a galaxy as the bar force ratio $Q_{\rm b}^c \equiv 4^{-1}\sum_{i=1}^4 Q_{\rm T,\it i}$, where $Q_{\rm T,\it i}$ indicates the maximum $Q_{\rm T}$ in the quadrant $i$. We used the superscript `c' to denote that it is obtained from the ratio map in the Cartesian coordinates in the manner of \citetalias{BB01}.

\subsection{Ratio Map in the Polar Coordinates} \label{chap3.3.1}

We have tried a different approach. We converted this ratio map $Q_{\rm T}(i,j)$ from the Cartesian coordinates to the polar coordinates, i.e., $Q_{\rm T}(r,\phi)$ as a function of $r$ and $\phi$. The ratio map can be described in the polar coordinates as 
\begin{eqnarray}\label{eq3}
Q_{\rm T}(r,\phi) = \frac{F_{\rm T}(r,\phi)}{\langle F_{\rm R}(r)\rangle}, 
\end{eqnarray}
where the radial force and tangential force are given by equation \ref{eq5} and equation  \ref{eq6}. It makes the complex patterns on the Cartesian grid dramatically simple and intuitive. 

In Figure \ref{fig3.3.1a}(c), we present the ratio map $Q_{\rm T}(r,\phi)$ for the same galaxy. The four wings of the butterfly pattern in the Cartesian grid are transformed to the four horizontal thick slabs, which are aligned with similar length ($r$) and width ($\phi$) in the $Q_{\rm T}(r,\phi)$ map. The four maxima in each quadrant (blue crosses in Figure \ref{fig3.3.1a}(b)) appear within each thick slab in the $Q_{\rm T}(r,\phi)$ map (blue crosses in Figure \ref{fig3.3.1a}(c)).

On the other hand, spiral arms are distinguished from the bar, appearing like ripples outside the bar (bottom row in Figure \ref{fig3.3.1a}(c)). This is one of the most important merits of this representation because the \citetalias{BB01} method often has a trouble that maximum points in quadrants are influenced by a bulge or by spiral arms \citep{2002Laurikainen_Salo, 2003Buta, 2005Buta, 2017Garcia}. \citet{2003Buta} developed a way to separate the bar strength from the spiral strength by analyzing the additional Fourier amplitude. Meanwhile, we disentangle the bar strength from the strength caused by a bulge and spiral arms by investigating the ratio map as a function of $r$ and $\phi$.

Besides, we noticed that the four thick slabs for a bar are warped when they are connected to the ripples of the spiral arms. The degree of warping may provide the information about the pitch angle of the spiral arms. In fact, the ratio map shows all kinds of galactic structures, a marble cake mixed with signatures from diverse components. 

In Figure \ref{fig3.3.1a}(d), we present the radial $Q_{\rm T}$ profile, which is the averaged $Q_{\rm T}$ over the azimuthal angle at each radius, 
\begin{eqnarray}
\langle Q_{\rm T}\rangle(r) = \frac{1}{2\pi}\int_{0}^{2\pi}Q_{\rm T}(r,\phi)d\phi.
\end{eqnarray}
This graph has been investigated in previous studies by \citet{2002Laurikainen_Salo} and \citet{2004aLaurikainen, 2004bLaurikainen}. This profile shows decreasing mean strengths from the center, followed by an increase to the maximum. The range from the center to the first minimum is the bulge-dominated region, denoted by the blue vertical dotted line. As the bar dominates, $\langle Q_{\rm T}\rangle(r)$ starts to increase and has a maximum peak, indicated by the red vertical solid line. We call this $r_{\rm Qb}$, the radius where $\langle Q_{\rm T}\rangle(r)$ reaches a maximum. This radius has been studied and used as one type of measurements for the bar length \citep{2002Laurikainen, 2002Laurikainen_Salo, 2004aLaurikainen, 2004bLaurikainen, 2016Diaz}.  

\begin{figure}[hbtp]
\centering
\includegraphics[bb = 30 660 155 790, width = 0.65\linewidth, clip = ]{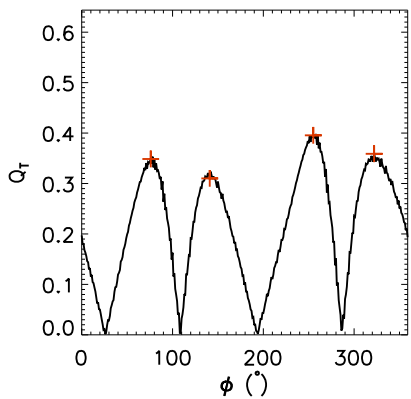}
\caption{Azimuthal profile $Q_{\rm T}(r_{\rm Qb},\phi)$ of UGC 06903 (Figure \ref{fig3.3.1a}) at $r_{\rm Qb}$, where the maximum of the radial profile $\langle Q_{\rm T}\rangle (r)$ is located. The butterfly pattern appears as four peaks (red crosses) along the azimuth $\phi$. \label{fig3.3.1b}}
\end{figure}

Now, we investigate the marble cake map of $Q_{\rm T}(r,\phi)$ in $(r, \phi)$ coordinates (Figure \ref{fig3.3.1a}(c)). First, we exclude the bulge region in the ratio map, shown as a black vertical stripe, in order to highlight the bar region. Then, we overlay $r_{\rm Qb}$ shown as the red solid line on the $Q_{\rm T}(r,\phi)$ map. It is located near the four maxima (blue crosses) used by \citetalias{BB01}. Cutting this marble cake of $Q_{\rm T}(r,\phi)$ at $r_{\rm Qb}$ yields the azimuthal profile of $Q_{\rm T}$ at $r_{\rm Qb}$, i.e., the variation of $Q_{\rm T}(r_{\rm Qb}, \phi)$ along the azimuth $\phi$. We plot the azimuthal profile of UGC 06903 in Figure \ref{fig3.3.1b} and find four maximum peaks, each corresponding to the maximum force ratio at each quadrant. We again overlay these four peaks on the deprojected image with red crosses in Figure \ref{fig3.3.1a}(a). They are located near the four maxima obtained from the \citetalias{BB01} procedures. 

We define, as the global bar strength measure, the bar force ratio of a galaxy in the polar coordinate representation as
\begin{eqnarray}\label{equ11}
Q_{\rm b} \equiv \frac{1}{m}\sum_{i=1}^m Q_{\rm T,\it i} 
\end{eqnarray}
where $Q_{\rm T,\it i}$ is the maximum value at each peak on the azimuthal profile $Q_{\rm T}(r_{\rm Qb},\phi)$ at $r_{\rm Qb}$, and $m$ is the number of the peaks. For typical bar structures, the value of $m$ is usually four, but in the presence of other structures, $m$ might be different from four. We use $Q_{\rm b}$ and $m$ values to classify barred galaxies in \S \ref{chap3.4.2}. 

\begin{figure}[hbtp]
\centering
\includegraphics[bb = 20 590 230 780, width = 0.8\linewidth, clip=]{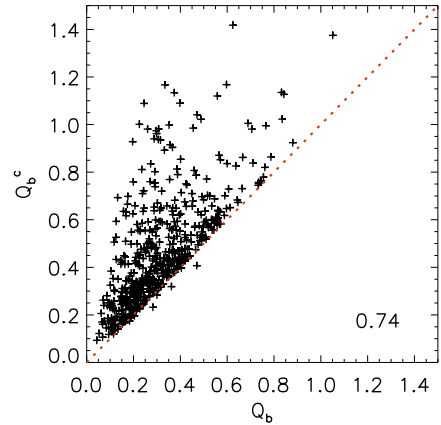}
\caption{Comparison of bar force ratio measurements in the polar coordinates $(Q_{\rm b})$ with that in the Cartesian coordinates $(Q_{\rm b}^c)$. The correlation coefficient is at the bottom right. The red dotted line denotes the track of a one-to-one correspondence. \label{fig3.3.1c}}
\end{figure}

In Figure \ref{fig3.3.1c}, we compare our bar force ratio value $Q_{\rm b}$ with $Q_{\rm b}^c$ measured by the \citetalias{BB01} method. They are quite correlated, but the values of $Q_{\rm b}^c$ are always greater compared to $Q_{\rm b}$. Theoretically, it is because the two bar force ratio measurements, $Q_{\rm b}^c$ and $Q_{\rm b}$, are different in that $Q_{\rm b}^c$ is the mean of global maxima of $Q_{\rm T}$ while $Q_{\rm b}$ is the mean of the maxima of $Q_{\rm T}$ at a given radius $r_{\rm Qb}$. However, we found that, in a large number of galaxies, the global maxima are contaminated by a bulge or spiral arms, as noted previously \citep{2002Laurikainen_Salo, 2003Buta, 2005Buta, 2017Garcia}. The local maximum derived at a certain radius helps the bar force ratio measurements not to be confused with other components, such as a bulge or spiral arms.

\subsection{Test on Mock Galaxies} \label{chap3.3.2}

\begin{figure*}[htbp]
\includegraphics[bb = 10 120 570 530, width = 0.98\linewidth, clip = ]{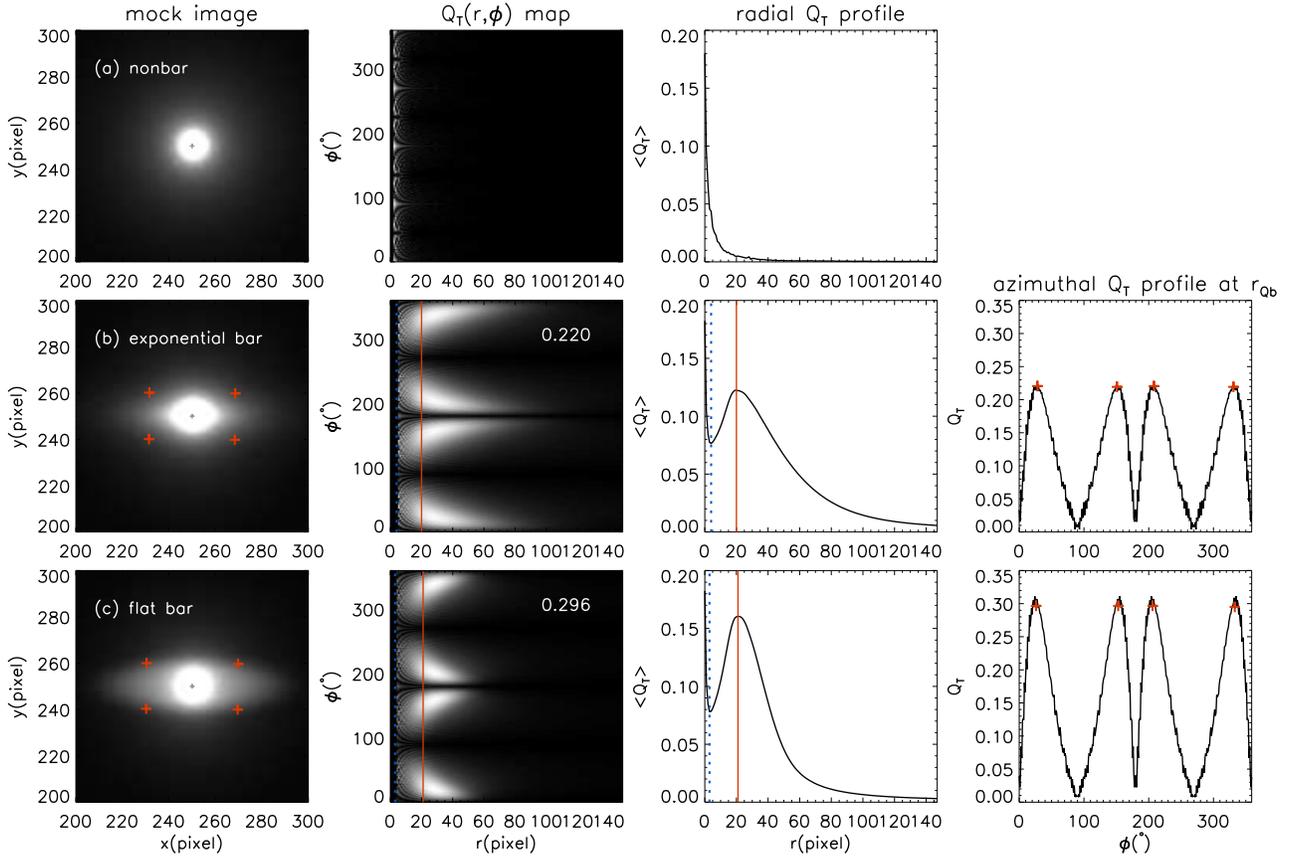}
\caption{Test on mock galaxies: (a) a nonbarred galaxy, (b) an exponential-barred galaxy, and (c) a flat-barred galaxy. All images contain a bulge with the bulge-to-total light ratio of 0.2 and a disk. From left to right, the mock image, ratio map $Q_{\rm T}(r,\phi)$, radial profile $\langle Q_{\rm T}\rangle(r)$, and azimuthal profile $Q_{\rm T}(r_{\rm Qb},\phi)$ at $r_{\rm Qb}$ are presented. Blue vertical dotted lines and red vertical solid lines present the minimum and maximum radius of $\langle Q_{\rm T} \rangle$, respectively. The inner region of the minimum radius of $\langle Q_{\rm T} \rangle$ is presented in black on the $Q_{\rm T}(r,\phi)$ map for barred galaxies to emphasize the bar pattern. Red crosses indicate peaks on the azimuthal profile at $r_{\rm Qb}$. We denote the bar force ratio $Q_{\rm b}$ on the top right in the second panel, the mean value of the peaks over the azimuthal angle at $r_{\rm Qb}$. \label{fig3.3.2}}
\end{figure*}

We constructed simple mock galaxies using GALFIT \citep{2010Peng} to test the properties of the $Q_{\rm T}(r,\phi)$ map before applying to observed galaxies. Mock galaxies are generated to have a bulge and a disk with or without a bar. The bulge is built with the S$\acute{e}$rsic index $n = 2$ and the disk component has an exponential radial profile. We considered two types of bars: a flat bar with a constant radial profile and an exponential bar with exponentially decreasing radial profile \citep{1985Elmegreen, 2015Kim}. Flat bars are more related to strongly barred galaxies. They are longer and stronger and have higher contrasts compared to exponential bars \citep{1985Elmegreen, 1989Elmegreen, 1986Baumgart, 1996Elmegreen, 1997Regan, 2015Kim, 2019Lee}. We have built flat bars with the S$\acute{e}$rsic index $n_{\rm bar}=0.25$ and exponential bars with $n_{\rm bar} = 0.85$, which are the mean values of $n_{\rm bar}$ from flat-barred and exponential-barred galaxies, respectively \citep{2015Kim}. We have fixed the axis ratio of the bar to 0.3.

Figure \ref{fig3.3.2} shows mock galaxies without a bar, with an exponential bar, and with a flat bar, from top to bottom. All have the same bulge-to-total ratio of 0.2. We present in Figure \ref{fig3.3.2} the mock image, $Q_{\rm T}(r,\phi)$ map, radial profile $\langle Q_{\rm T} \rangle(r)$, and azimuthal profile $Q_{\rm T}(r_{\rm Qb}, \phi)$ at $r_{\rm Qb}$ from left to right for each example. We confirm that the butterfly patterns are converted to four aligned thick slabs on the $Q_{\rm T}(r,\phi)$ map for both flat-barred and exponential-barred galaxies (second panels from left in Figures \ref{fig3.3.2}(b) and (c)). On the other hand, a mock galaxy without a bar shows a totally different pattern on the $Q_{\rm T}(r,\phi)$ map (second panel from left in Figure \ref{fig3.3.2}(a)). It shows a non-vanishing $Q_{\rm T}$ pattern only near the center, where digitizing errors produce artificial transverse forcing. We subtract this pattern caused by a bulge on other $Q_{\rm T}(r,\phi)$ maps throughout this paper, to highlight the patterns by bars.

The radial profile $\langle Q_{\rm T}\rangle(r)$ of a nonbarred galaxy is also quite different from those of both types of barred galaxies (third panels from left in Figure \ref{fig3.3.2}). The radial profile gradually decreases in the nonbarred galaxy (Figure \ref{fig3.3.2}(a)). However, when the mock galaxies have a bar, the radial profiles decrease from the center to reach a minimum, then followed by a maximum (Figures \ref{fig3.3.2}(b) and (c)). We regard the range from the center to the minimum as the bulge-dominated region and the region around the maximum as the bar-dominated region. Flat bars have a larger maximum value than exponential bars. 

In the fourth panels from left, we display the azimuthal profile of $Q_{\rm T}(r_{\rm Qb},\phi)$ at $r_{\rm Qb}$. Four wings of the butterfly pattern for a bar appear as four peaks on the azimuthal profiles (Figures \ref{fig3.3.2}(b) and (c)). We overlay the four peaks as red crosses on the mock images in the first panels and display the bar force ratio of a galaxy $Q_{\rm b}$ at the top right in the second panels. The flat-barred galaxy has a larger bar force ratio than the exponential-barred galaxy when they have the same bulge and disk. For the nonbarred mock galaxy (Figure \ref{fig3.3.2}(a)), we cannot display the azimuthal profile at $r_{\rm Qb}$, because it does not have any maximum peak on the radial profile, except for the center.

\section{Classification} \label{chap3.4}

\subsection{Two Types of the Radial Profile}\label{chap3.4.1}

\begin{figure*}[htbp]
\includegraphics[bb = 20 430 560 800, width = 0.98\linewidth, clip = ]{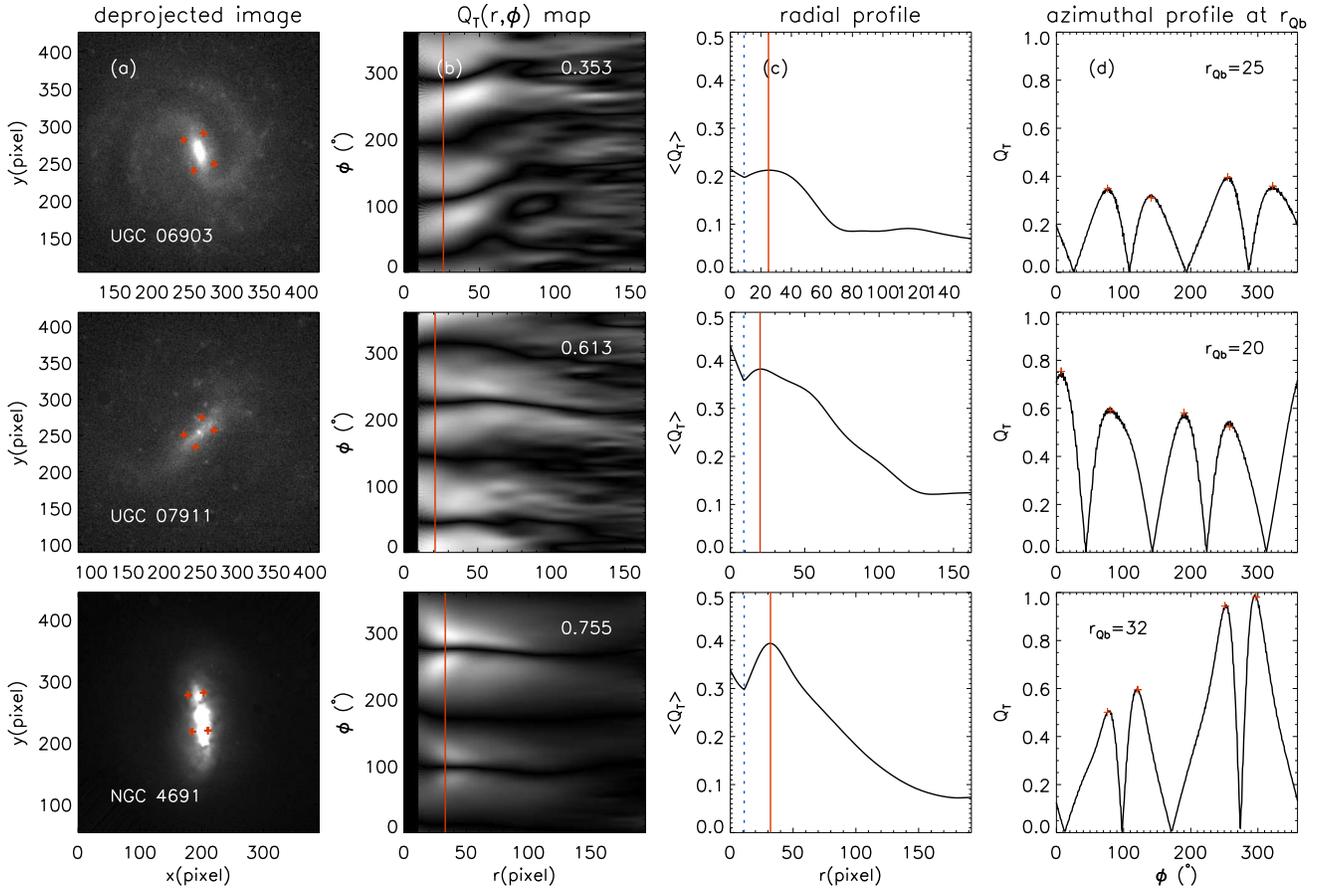}
\caption{Example type M galaxies, which have a local maximum peak on the radial profile. For each galaxy, we present (a) the deprojected image, (b) $Q_{\rm T}(r,\phi)$ map, and (c) the radial and (d) azimuthal profile of $Q_{\rm T}$. Blue vertical dotted lines in the panel (c) indicate the end of the bulge-dominated region. Red vertical solid lines present $r_{\rm Qb}$, corresponding to the bar-dominated region. Red crosses indicate peaks on the azimuthal profile at $r_{\rm Qb}$. The bar strength $Q_{\rm b}$ is denoted on the top right in the panel (b). \label{fig3.4.1a}}
\end{figure*}

Now we analyze $Q_{\rm T}(r,\phi)$ maps for our 884 sample galaxies. The first step in automatically finding the bar signature is to determine $r_{\rm Qb}$ on the radial profile. In Figure \ref{fig3.4.1a}, we display representative examples that have similar features shown in the analysis of mock barred galaxies (Figure \ref{fig3.3.2}). Although the shapes of radial profiles are not the same, all have a minimum point followed by a local maximum peak (Figure \ref{fig3.4.1a}(c)). We measure $r_{\rm Qb}$ at the local maximum peak in this case, denoted by the red vertical solid line.  

\begin{figure*}[htbp]
\includegraphics[bb = 20 430 560 800, width = 0.98\linewidth, clip = ]{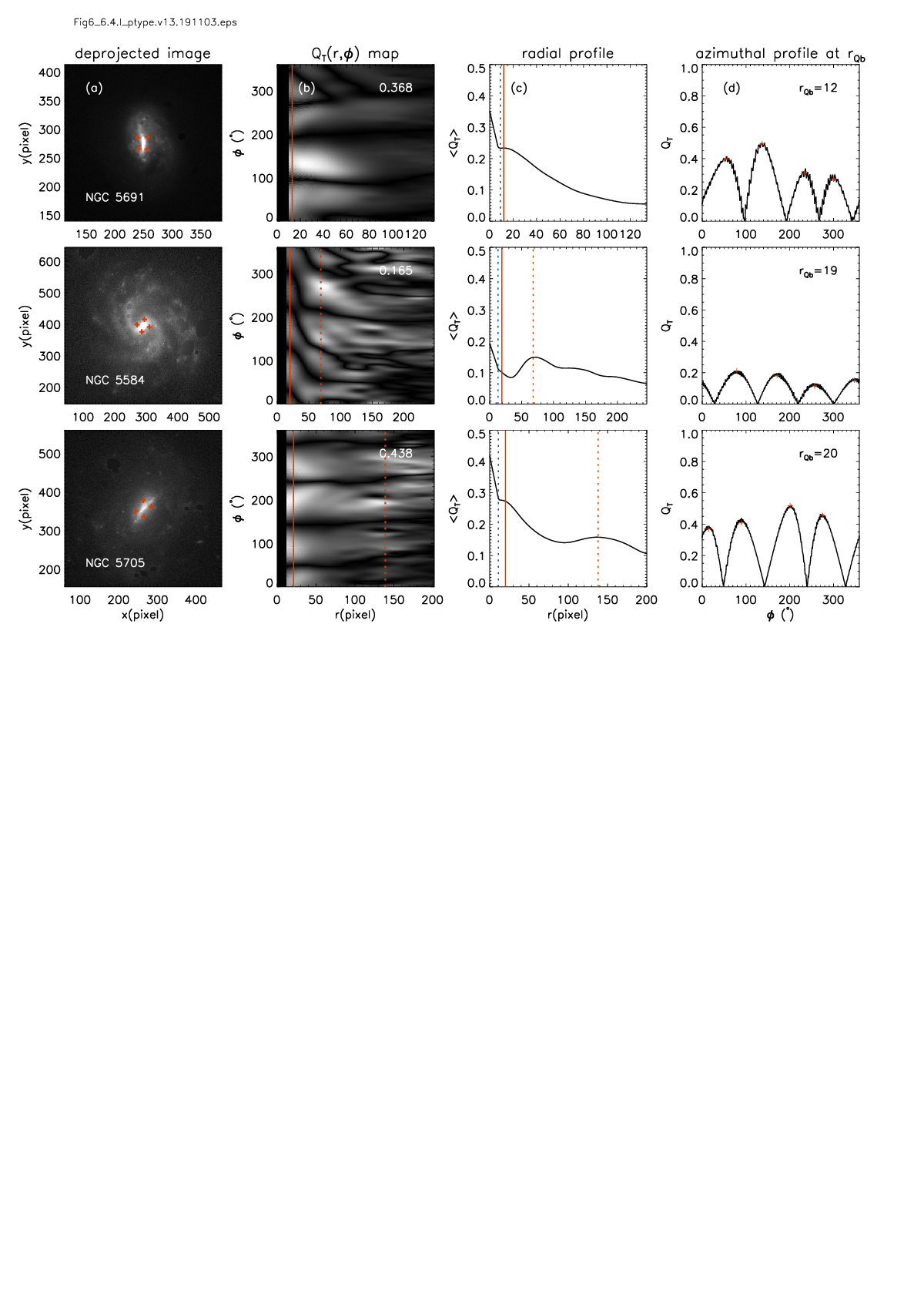}
\caption{Example type P galaxies, which has a plateau in the bar region in the radial profile. For each galaxy, we present (a) the deprojected image, (b) $Q_{\rm T}(r,\phi)$ map, and (c) the radial and (d) azimuthal profile of $Q_{\rm T}$ . As in Figure \ref{fig3.4.1a}, blue dotted lines, red solid lines, and red crosses indicate the limit of the bulge-dominated region, $r_{\rm Qb}$, and four peaks of a bar at $r_{\rm Qb}$, respectively. In the cases that the radial profile has not only a plateau but also a local maximum peak, we mark the radius for a local maximum with the red vertical dotted line. The bar strength $Q_{\rm b}$ is displayed on the top right in the panel (b). \label{fig3.4.1b}}
\end{figure*}

On the other hand, we find another type that does not have a local maximum in the bar region (Figure \ref{fig3.4.1b}(c)). Their radial profiles $\langle Q_{\rm T} \rangle (r)$ have a kind of plateau instead of a local maximum. When we cut the marble cake of $Q_{\rm T}(r,\phi)$ at the inflection radius where a plateau resides, 
\begin{eqnarray}
\frac{d^2(\langle Q_{\rm T}\rangle(r))}{dr^2} = 0,
\end{eqnarray}
we also find four peaks, the signature for a bar on the azimuthal profile, as shown in Figure \ref{fig3.4.1b}(d). We present the four peaks with red crosses on the deprojected images and confirm that they are located around the four corners of the bar (Figure \ref{fig3.4.1b}(a)). Hence, in such case, we define $r_{\rm Qb}$ as the radius of inflection. We indicate $r_{\rm Qb}$ by red vertical solid lines in Figures \ref{fig3.4.1b}(b) and (c). 

We investigate images for sample galaxies of these two types by visual inspections, and notice that more developed barred galaxies with grand-design spiral arms or rings usually have a maximum peak in their radial profiles. On the other hand, galaxies with a plateau in their radial profiles rarely show prominent structures, except for bars. We suspect that the different features in the radial profile might be related to the stages in bar evolution. We further investigate these two types in comparison with the analysis for the numerical simulation from \citet{2019Seo} and discuss it in \S \ref{chap3.6.2}. Here, we roughly classify galaxies into two types on the radial profiles $\langle Q_{\rm T}\rangle(r)$: one that has a local maximum peak and the other that has a plateau with an inflection point. We call the former `type M', standing for `maximum peak', and the latter `type P', signifying to the `plateau'. 

\begin{figure*}[hbtp]
\centering
\includegraphics[bb = 40 660 410 810, width = 0.95\linewidth, clip = ]{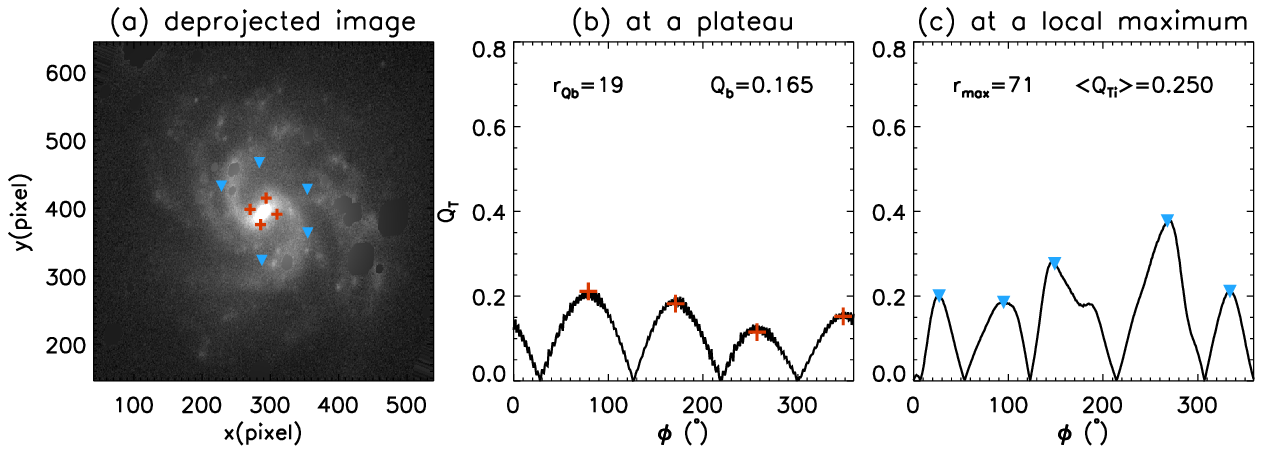}
\caption{An example galaxy NGC 5584 that has both a plateau and a maximum peak on the radial profile. We present the deprojected image in (a) and azimuthal profiles at a plateau in (b) and a local maximum peak in (c). We display the radius of the plateau and the local maximum at the top left and the mean strength value of peaks at the top right in each panel. Then each peak on the azimuthal profiles is overlaid on the deprojected image by red crosses and skyblue triangles. The four peaks at the plateau are located near the bar and five peaks at the maximum radius in the middle of spiral arms.\label{fig3.4.1c}}
\end{figure*}

Naturally, there are other diverse cases in real galaxies. NGC 5584 and NGC 5705 have not only a plateau but also a local maximum in the radial profile, which are denoted by red solid and dotted lines, respectively, in Figures \ref{fig3.4.1b}(b) and (c). Similarly, there are some galaxies that have two local maximum peaks. For these cases, we investigate all azimuthal profiles for a plateau or local maximum peaks. For example, in Figure \ref{fig3.4.1c}, we compare the azimuthal profiles of NGC 5584 at both radii of a plateau and of a local maximum. The profile at the plateau has four peaks for a bar signature (Figure \ref{fig3.4.1c}(b)), whereas that at the local maximum shows five peaks (Figure \ref{fig3.4.1c}(c)). We overlay each peak on the deprojected image by red crosses and skyblue triangles in Figure \ref{fig3.4.1c}(a). We confirm that the five peaks at the local maximum are caused by spiral arms. We calculate the mean value $\langle Q_{\rm T, \it i}\rangle$ of peaks on the azimuthal profile at each radius (equation \ref{equ11}) and denote it at the top right of each panel in Figure \ref{fig3.4.1c}. Even though the $\langle Q_{\rm T, \it i}\rangle$ value at the local maximum peak is larger than that at the plateau, we conclude that the large value is caused by spiral arms, not by the bar. Therefore, we preferentially adopt $r_{\rm Qb}$ where the azimuthal profile shows the bar signature, i.e., four peaks. 

\subsection{Two Classes on the Azimuthal Profile: Bar versus Nonbar} \label{chap3.4.2}

As discussed above, we can distinguish bars from spiral arms on the $Q_{\rm T}(r,\phi)$ map: bar structures appear as four thick horizontal slabs, whereas spiral arms as ripples. It becomes more apparent when we investigate the azimuthal profiles at their $r_{\rm Qb}$, as shown in Figure \ref{fig3.4.1c}. Bars have exactly four peaks while spiral arms usually have more than four peaks. It is the most essential criterion in distinguishing barred galaxies from nonbarred galaxies. We further constrain $Q_{\rm b}$ to be greater than a certain threshold ($Q_{\rm b} \ge 0.15$) in order to exclude very weak ovals or elongated bulges. Some of the elongated bulges in the deprojected images are artifacts due to simple deprojection of spherical bulges. 

We classify a galaxy as a barred galaxy if it has (1) four peaks in the azimuthal profile of $Q_{\rm T}$ at $r_{\rm Qb}$ and (2) the bar strength $Q_{\rm b}$ above 0.15. Using these criteria, we classified 468 spirals as barred galaxies (53\%) and 416 spirals as nonbarred galaxies (47\%). This bar fraction is close to the bar fraction (60\%) including both SBs and SABs by the classical visual inspection \citep{1973Nilson, 1987Sandage, RC3, 2015Buta, Ann15}, but is higher than that by previous automated classifications such as the ellipse fitting method ($\sim45\%$) or Fourier analysis ($\sim40\%$) \citep{2002Laurikainen_Salo, 2007Marinova, 2007Reese, 2008Barazza, 2009Aguerri, 2010Marinova, 2019Lee}.

\begin{figure}[htbp]
\includegraphics[bb = 210 530 460 680, width = 0.98\linewidth, clip=]{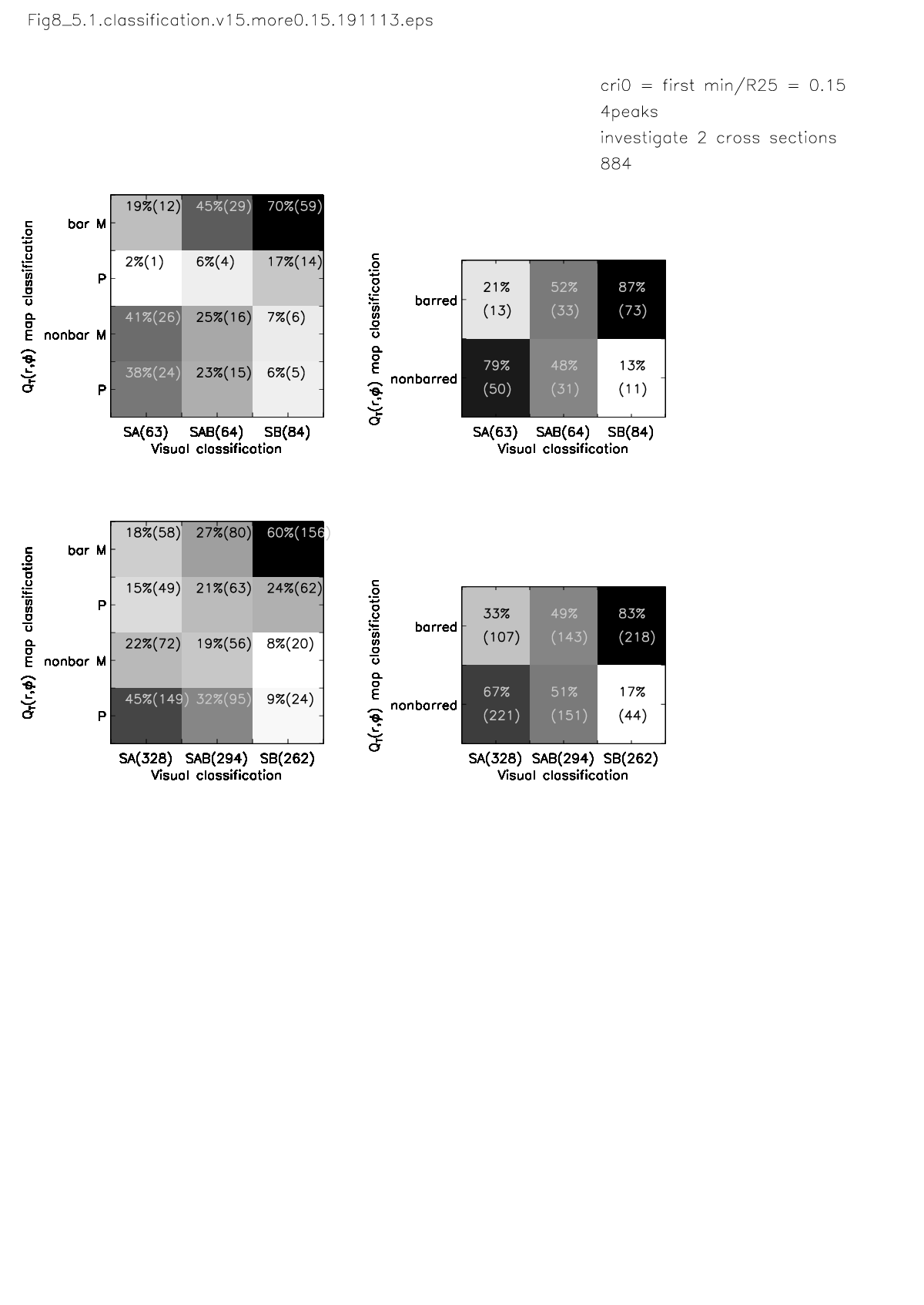}
\caption{Comparison between our automated classification utilizing $Q_{\rm T}(r,\phi)$ map and the visual classification from RC3 \citep{RC3} and \citetalias{Ann15} catalogs. The numbers mean matched percentages of the automated classification (ordinate) for the visual classification (abscissa). The larger number and darker shade indicate better agreement. \label{fig3.4.2a}}
\end{figure}

\begin{figure*}[htbp]
\includegraphics[bb = 10 120 580 780, width = 0.98\linewidth, clip = ]{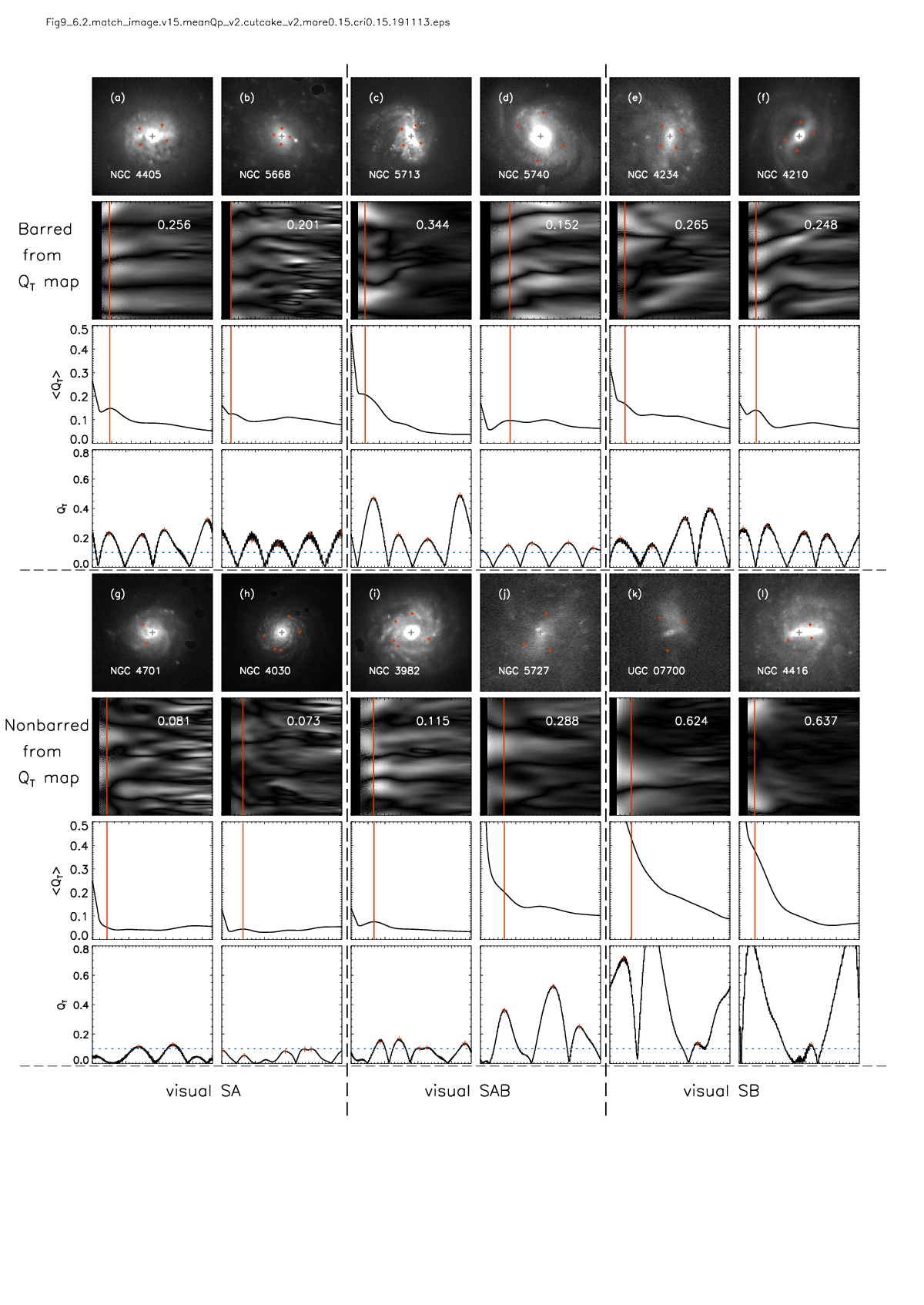}
\caption{Representative example galaxies for comparison between automated classification by $Q_{\rm T}(r,\phi)$ map analysis and the visual classification from RC3 \citep{RC3} and \citetalias{Ann15} catalogs. The upper row shows automatically classified SB galaxies and the bottom row SA galaxies. The visual classification is presented as SA, SAB, and SB in the abscissa from left to right. Each galaxy is shown with the deprojected image, $Q_{\rm T}(r,\phi)$ map, and the radial and azimuthal profile from top to bottom. \label{fig3.4.2b}}
\end{figure*}

\subsection{Comparison with the Visual Classification} \label{chap3.4.3}

We compare our new automated classification method against the visual inspection. For comparison, we used 63 SA, 64 SAB, and 84 SB galaxy `concordance' sample in \citetalias{2019Lee}. It is the sample of galaxies classified as the same class by two independent visual inspections, RC3 \citep{RC3} and \citetalias{Ann15}. Figure \ref{fig3.4.2a} shows the consistency between our automated classification (ordinate) and the visual classification (abscissa). Our method shows good agreements with the visual classification by 79\% and 87\% for the SA and SB classes, respectively. However, our method classifies about half of visually classified SABs as SBs and the other half as SAs, which is understandable because SABs by definition show weak bar features. 

We display the example of the matched and unmatched galaxies in Figure \ref{fig3.4.2b}. The galaxies in the upper row are barred galaxies (SBs) determined by our $Q_{\rm T}(r,\phi)$ map analysis and those in the bottom row nonbarred galaxies (SAs). The visual classification is displayed as SA, SAB, and SB in the abscissa from left to right. For each galaxy, we present the deprojected image, $Q_{\rm T}(r,\phi)$ map, and the radial and azimuthal profile at $r_{\rm Qb}$ from top to bottom. 

We confirm that the azimuthal profiles of the upper row are definitely different from those of the bottom row, although they are unmatched against the visual classification. Whereas the azimuthal profiles on the upper row show exactly four peaks, those on the bottom row show more or less than four peaks. The shapes of peaks look different as well: the peaks of the upper row look regular and similar to each other, whereas those of the bottom row look very irregular. In fact, we found four additional barred galaxies that the previous visual inspections did not find, as shown in Figure \ref{fig3.4.2b}(a), and a galaxy that has oval structures, as shown in Figure \ref{fig3.4.2b}(b), among the unmatched galaxies (13 galaxies) against visually determined nonbarred galaxies. Accordingly, our misjudgement against nonbarred galaxies becomes 13\% (eight galaxies) in practice, not 21\%.

The main factor to cause the confusion is the deprojection process. If the inclination is underestimated, the less-deprojected disk may be confused as a bar. Overestimated inclination, on the other hand, stretches the disk along the minor axis. Therefore, we need to develop a more careful deprojection process to deal correctly with the spherical component, i.e., bulge, in order to increase the accuracy of the classification.  

We missed 11 visually determined SB galaxies (13\%). One of them is disturbed by the residuals after masking bright clumpy sources, and three have $Q_{\rm b}$ values lower than 0.15 because of a large bulge, which will be discussed in \S\ref{chap3.6.1}. Meanwhile, we found interesting features in the rest of them: they have asymmetric structures. In Figure \ref{fig3.4.2b}(l), NGC 4416 has two nuclei within its bar structure. The peak deviates significantly from the centroid. We denote the peak by a gray cross on the deprojected image. If we use the centroid as the center instead of the peak for this galaxy, we find the normal signature for a bar on the $Q_{\rm T}(r,\phi)$ map and the azimuthal profile. We found six other asymmetric bars like UGC $07700$ shown in Figure \ref{fig3.4.2b}(k). They usually do not have any bulge but have small bright sources in the elongated bar-like structures. The nucleus is displaced from the center of the bar. These are galaxies with unusual dynamics. 

Lastly, we investigate how the inclination of a galaxy influences the accuracy of our automated classification. We excluded galaxies inclined more than 60$^\circ$ and constructed face-on galaxy images through the deprojection. We found that inadequate deprojection is the main factor causing mistakes in our automated classification as mentioned before.

We then investigate the agreement and disagreement rate of our automated classification against the visual classification by restricting the sample galaxies to inclinations less than 60$^\circ$, 45$^\circ$, and 30$^\circ$. In Table \ref{Table3.4.2}, we organize the results both for the concordance sample and the whole sample of galaxies. We find that the effect of the inclination on the classification accuracy appears in SAs, but not in SBs. For the whole sample, the agreement rates are improved by 5\% and 4\% as the inclination restriction is lowered from 60$^\circ$ to 45$^\circ$ and 30$^\circ$. This is because the more a galaxy is inclined, the less accurately the inclination is measured. A spherical bulge makes the inclination underestimated in high inclination galaxies. Consequently, the less-deprojected disk is easily confused as a bar, as discussed before. For the concordance sample, the agreement rate for 45$^\circ$ restriction is reduced compared to that for 60$^\circ$ restriction, but it is not statistically meaningful because the corresponding sample is too small. 

Strongly barred galaxies seem to be less affected by the galaxy inclination. In the whole sample, the agreement rate increases by 3\% as the inclination restriction is lowered from 60$^\circ$ to 45$^\circ$, but barred galaxies are mainly misjudged for other reasons such as the residuals after masking, asymmetric bars, or lower $Q_{\rm b}$ than 0.15 as discussed before, not by the inclination or deprojection. We do not find elongated bulges by the deprojection in the misjudged SBs. They are already excluded from barred galaxies because they have $Q_{\rm b}$ lower than 0.15. We do not find any variation in the agreement rate between 45$^\circ$ and 30$^\circ$ restriction. 

\begin{table*}[htb]
 \caption{The Effect of the Inclination of Galaxy on the Agreement and Disagreement between the Visual Classification and the $Q_{\rm T}$ Map Classification \label{Table3.4.2}}
\begin{center}
\begin{tabular}{l|l|l|rr|rr|rr}
\hline 
 & \multicolumn{2}{c|}{Inclination restriction} & 
 \multicolumn{2}{c|}{60$^\circ$} & 
 \multicolumn{2}{c|}{45$^\circ$} & 
 \multicolumn{2}{c}{30$^\circ$} \\ 
\cline{2-9} 
 & \multicolumn{2}{c|}{Classification method} & 
 \multicolumn{2}{c|}{Sample} & 
 \multicolumn{2}{c|}{Sample} & 
 \multicolumn{2}{c}{Sample} \\ 
\cline{2-9} 
{ } & {Visual inspection} & {$Q_{\rm T}$ map} & {con\footnote{It means the concordance sample that are classified as the same class by two independent visual inspections, RC3 and Ann15.} (211)} & {all\footnote{It means the whole sample.} (884)} & {con (143)} & {all (560)} & {con (75)} & {all (271)} \\ 
\hline
          & SA  & SA  & 79\% (50) & 67\% (221) & 76\% (32) & 72\% (148) & 83\% (20) & 76\% (77)  \\
Agreement & SAB & SB & 52\% (33) & 49\% (143) & 56\% (23) & 51\% (94) &  53\% (8) & 56\% (45) \\
          & SB  & SB  & 87\% (73) & 83\% (218) & 88\% (53) & 86\% (148) & 89\% (32) & 86\% (77) \\ 
\hline
             & SA  & SB  & 21\% (13) & 33\% (107) & 24\% (10) & 28\% (57) &  17\% (4) & 24\% (24) \\
Disagreement & SAB & SA  & 48\% (31) & 51\% (151) & 44\% (18) & 49\% (89) &  47\% (7) & 44\% (35) \\
             & SB  & SA  & 13\% (11) & 17\% (44) &  12\% (7) & 14\% (24) &  11\% (4) & 14\% (13) \\
\hline
\multicolumn{9}{l}{\footnotesize The numbers in the parentheses indicate the number of galaxies.}\\
\end{tabular}
\end{center}
\end{table*}


\subsection{SB versus SAB}\label{chap3.4.3}

We can speculate that weak bars (SABs) will be distinguished from strong bars (SBs) in terms of the bar strength. However, it is not straightforward because the bar strengths of strong bars and weak bars are distributed over a wide range \citep{BB01, 2001Block, 2002Laurikainen_Salo, 2004Buta, 2019Cuomo}.

\begin{figure}[htbp]
\includegraphics[bb = 20 630 170 770, width = 0.9\linewidth, clip=]{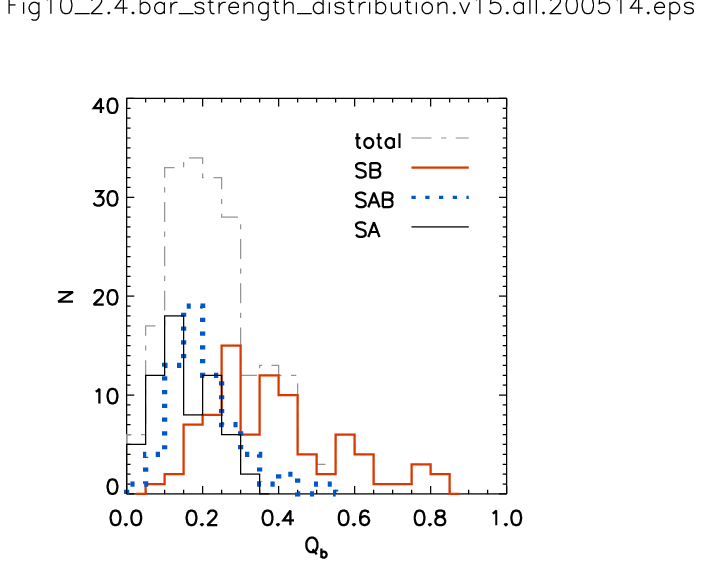}
\caption{Distribution of the bar force ratio $Q_{\rm b}$ for visually classified SA, SAB, and SB galaxies. The gray dot-dashed line indicates total galaxies. We denote SA, SAB, and SB galaxies by thin black solid line, thick blue dotted line, and thick red solid line, respectively.\label{fig3.4.3a}}
\end{figure}

Figure \ref{fig3.4.3a} displays the distribution of the bar force ratio $Q_{\rm b}$ for visually determined SA, SAB, and SB glaxies. The force ratios for SA galaxies are defined by the mean values of the $Q_{\rm T}$ peaks along the azimuthal profile at $r_{\rm Qb}$ as well. The only difference from the barred galaxies is the number of peaks. 

SB galaxies are known to have larger $Q_{\rm b}$ than SAB galaxies \citep{BB01, 2004Buta, 2005Buta, 2007Buta}: its distribution is statistically distinguishable from that of SABs. However, their values of the bar force ratio cover a wide range from $Q_{\rm b}=0.1$ to 0.8 as shown in Figure \ref{fig3.4.3a}, which is consistent with earlier studies \citep{BB01, 2001Block, 2002Laurikainen_Salo, 2004Buta}. On the other hand, the bar force ratios of SAB galaxies are distributed in a narrower range than those of SB galaxies, and are similar to the distribution of force ratios for SAs rather than those for SBs (Figure \ref{fig3.4.3a}), as shown in previous studies \citep{2000Abraham, 2004aLaurikainen, 2017Garcia}. Nevertheless, it is hard to separate individual SAB galaxies from SB galaxies based directly on $Q_{\rm b}$ because they overlap in the distribution of bar force ratio. It could result from the fact that weak ovals may have a significant amount of mass.

\begin{figure}[htbp]
\includegraphics[bb = 30 600 230 780, width = 0.98\linewidth, clip=]{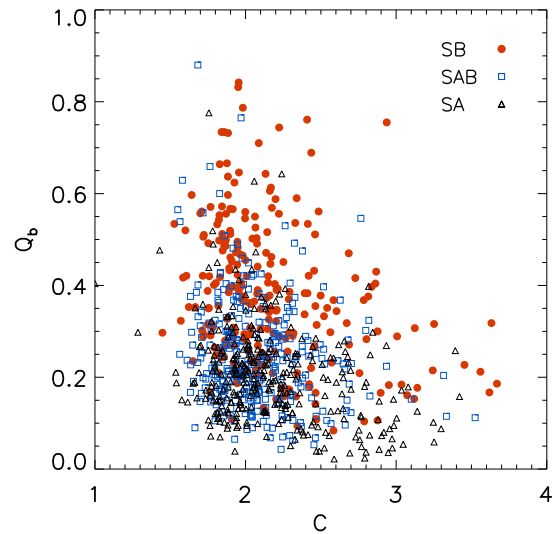}
\caption{Distribution of visually classified SA (black triangle), SAB(blue square), and SB (red solid circle) galaxies in the diagram of the light concentration ($C \equiv R_{90}/R_{50}$) versus the bar force ratio $Q_{\rm b}$. \label{fig3.4.3b}}
\end{figure}

\citet{2000Abraham} showed the separation between SB and SAB galaxies on the diagram of central concentration versus bar strength for a small sample of 56 galaxies. They defined the bar strength as
\begin{eqnarray}\label{equ3.4.3}
S_{\rm bar} = \frac{2}{\pi}[{\rm arctan}(b/a)^{-1/2}_{\rm bar}-{\rm arctan}(b/a)^{+1/2}_{\rm bar}]
\end{eqnarray}
 where a and b are the major and minor axis of the bar, respectively. In Figure \ref{fig3.4.3b}, we plot visually classified SB (red solid circle), SAB (blue square), and SA (black triangle) galaxies in the diagram of the light concentration ($C \equiv R_{\rm 90}/R_{\rm 50}$) versus the bar force ratio ($Q_{\rm b}$). We find similar distributions with \citet{2000Abraham} who showed a broad anti-correlation between the light concentration and the bar strength, and a skewed distribution of SAB toward lower $Q_{\rm b}$ compared to SB galaxies. However, we do not confirm the separation between SB and SAB galaxies. Even when we have used the bar strength (Equation \ref{equ3.4.3}) defined by \citet{2000Abraham}, the result was not different.

\begin{figure}[htbp]
\includegraphics[bb = 0 610 240 810, width = 0.98\linewidth, clip=]{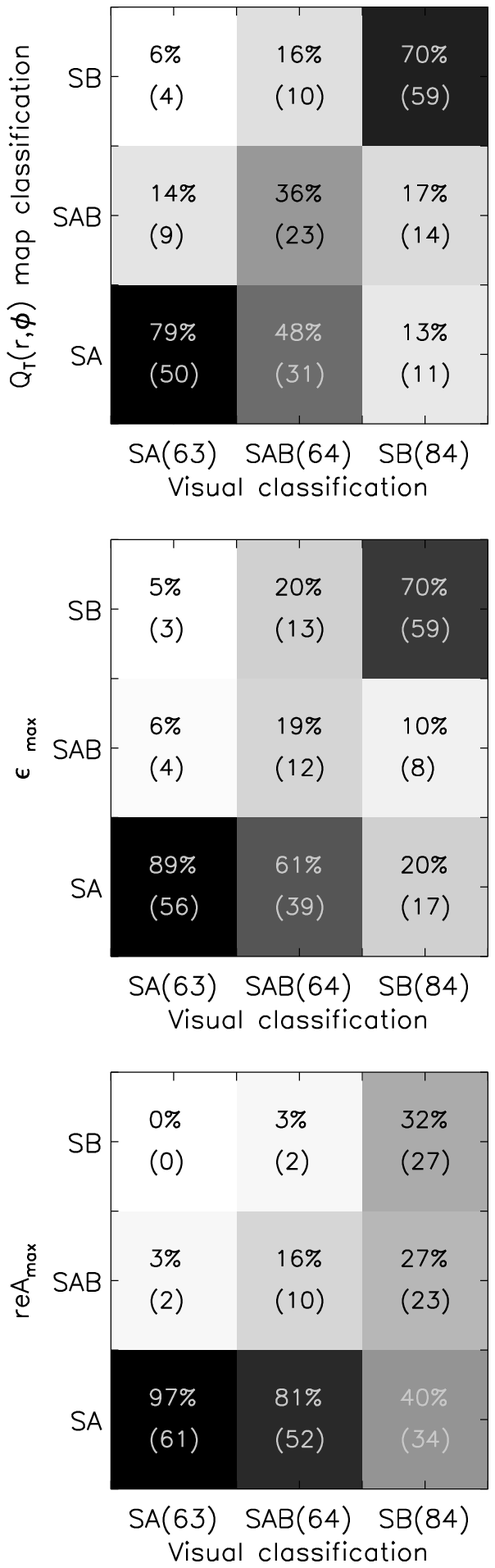}
\caption{Comparison between our classification using bar force ratio $Q_{\rm b}$ and the visual classification from RC3 \citep{RC3} and \citetalias{Ann15} catalogs. The numbers show matched percentages of the automated classification (ordinate) for the visual classification (abscissa). \label{fig3.4.3c}}
\end{figure}

Nevertheless, we attempted to classify barred galaxies into SB and SAB galaxies based on the value of $Q_{\rm b}$ alone. We set the criterion $0.15 \le Q_{\rm b} < 0.25$ for SAB, which is the range that SAB galaxies are dominant as shown in Figure \ref{fig3.4.3a}. We categorized galaxies with $Q_{\rm b} \ge 0.25$ as SBs. We classified the concordance sample galaxies used in Figure \ref{fig3.4.2a} according to this criterion. We present the consistency against the visual classification in Figure \ref{fig3.4.3c} as in Figure \ref{fig3.4.2a}. This criterion classifies 36\% of visually selected SAB galaxies as SABs. Although it is somewhat lower compared to SA or SB classes, it is the highest agreement for SABs when compared with other automated classification methods such as the ellipse fitting method shown in \citetalias{2019Lee}. 

When we investigate the unmatched SABs, 16\% are classified as SB because of $Q_{\rm b}$ larger than 0.25, whereas 9\% are classified as SAs due to $Q_{\rm b}$ smaller than 0.15. However, 39\% of SABs do not satisfy the criterion of four peaks in the azimuthal profile of $r_{\rm Qb}$. It shows that many of SAB galaxies have characteristics similar to those of SA galaxies in terms of not only the bar strength but also of the pattern on the ratio map. This supports the previous speculations that SAB galaxies have a similar origin of the force ratio as SA galaxies \citep{2004bLaurikainen} or that SAB galaxies would be the extension of SA galaxies \citep{2000Abraham}. It may also be that SAB galaxies are more heterogeneous than SB galaxies; visual SAB classification can refer to ovals as well as weak bars. In some galaxies, both an oval and a regular bar co-exist. 

\section{Bar Properties} \label{chap3.5}

\subsection{Bar Strength and Length} \label{chap3.5.1}

The $Q_{\rm T}(r,\phi)$ map analysis not only determines whether a bar is present or not but also provides important properties such as the bar strength $Q_{\rm b}$ and the bar length $r_{\rm Qb}$. The photometric bar strength $Q_{\rm b}$ is tightly correlated to the kinematic bar strength $Q_{\rm kin}$ measured from the stellar velocity field \citep{2015Seidel}. Moreover, it is well correlated to other bar strength measurements such as the bar ellipticity $\epsilon_{\rm bar}$ \citep{2001Block, 2002Laurikainen, 2002Laurikainen_Salo, 2016Diaz} and the normalized Fourier amplitude of $m = 2$ over $m = 0$ components, $A_2$ \citep{2002Laurikainen, 2016Diaz, 2019Cuomo}. The radius $r_{\rm Qb}$ has also been reported to be correlated to the bar length $r_{\rm vis}$ estimated by visual inspection \citep{2002Laurikainen, 2016Diaz} and $r_{\rm A2}$ by the Fourier analysis \citep{2004aLaurikainen}. 

\begin{figure*}[htbp]
\includegraphics[bb = 10 420 580 800, width = 0.99\linewidth, clip=]{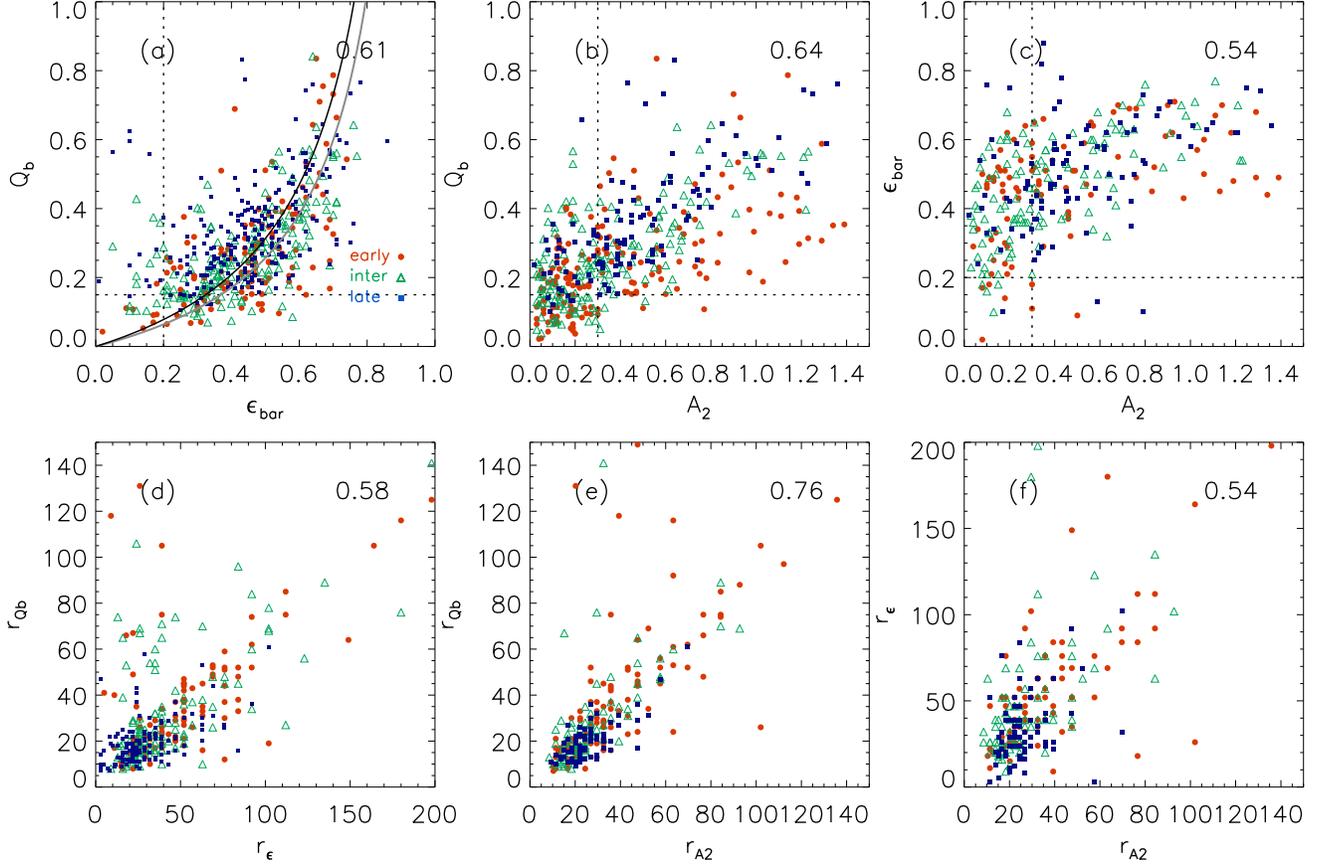}
\caption{Comparison of bar strength ({\it upper row}) and length ({\it bottom row}) measurements between the $Q_{\rm T}(r,\phi)$ map analysis, ellipse fitting method, and Fourier analysis. In the upper row, we overplot the threshold $Q_b = 0.15$, $\epsilon_{\rm bar} = 0.2$ , and $A_2 = 0.3$ for the bar detection of the three classification methods by the dotted lines. And the galaxies under the threshold are excluded in the bottom row. Galaxies are separated into early-type (red solid circle), intermediate-type (green open triangle), and late-type spirals (blue solid square). The correlation coefficients are presented at the top right of each panel. In the panel (a), we overplot the analytic relation between $\epsilon_{\rm bar}$ and $Q_{\rm b}$ \citep{2012Kim}, which depends on the bar mass fraction $f_{\rm b}$. The black and gray lines indicate the relations when $f_{\rm b}=0.5$ and 0.4 for a homogeneous bar ($n = 0$), respectively. \label{fig3.5.1}}
\end{figure*}

In Figure \ref{fig3.5.1}, we compare the various bar strength and length measurements by the $Q_{\rm T}(r,\phi)$ map analysis, ellipse fitting method, and Fourier analysis. In the ellipse fitting method, a bar is defined by the criteria on four parameters: the transition between a bar and a disk with $\Delta \epsilon_{\rm tra} \ge 0.1$ and $\Delta \rm PA_{\rm tra} \ge 5^\circ$, the ellipticity threshold of $\epsilon_{\rm bar} \ge 0.25$, and constant position angle $\Delta \rm PA_{\rm bar}$ within $\pm 10^\circ$ \citep{2004Jogee, 2019Lee}. For the Fourier analysis, the bar region should have the normalized Fourier intensity amplitudes larger than $A_2 = 0.3$ along with the $m = 2$ and $m = 4$ phases maintained constant \citep{1990Ohta, 2002Laurikainen, 2019Lee}. Although the definition of the normalized Fourier amplitude $A_2$ is somewhat different depending on the studies \citep{2013Athanassoula, 2019Seo}, we use $A_2$ defined as $I_2/I_0$ throughout this paper, which is described by
\begin{eqnarray}
A_2 \equiv \frac{I_{\rm 2}(r)}{I_{\rm 0}(r)} = \frac{[a_{\rm 2}(r)^2+b_{\rm 2}(r)^2]^{1/2}}{a_{\rm 0}(r)/2}
\end{eqnarray}
where $a_0$, $a_2$, and $b_2$ are the Fourier coefficients. \citep{1990Ohta, 2019Lee}. The bar strength is defined by the maximum $\epsilon$ or $A_2$ in the bar region \citep{1995Martin, 2002Laurikainen, 2002Laurikainen_Salo}, and the bar length $r_{\epsilon}$ or $r_{A2}$ by the radius where the maximum of $\epsilon$ or $A_2$ occurs, which can be measured reliably, but is not the full length of the bar \citep{1995Wozniak, 2002Laurikainen, 2002Laurikainen_Salo, 2016Diaz}. We used the $i-$band deprojected images for all methods in order to compare them in the same manner. For the ellipse fitting method, we used the criteria of $\epsilon_{\rm bar} \ge 0.2$ and $\Delta \rm PA_{\rm tra} \ge 2^\circ$, which are the best criteria for the $i-$band deprojected images that yields the highest agreement with the visual classification \citepalias{2019Lee}. 

Figure \ref{fig3.5.1} shows comparatively good correlations between different measurements of bar strength (upper row) and length (bottom row), as reported in the literature \citep{2001Block, 2002Laurikainen, 2002Laurikainen_Salo, 2016Diaz}, although the scatters are not small. We find that the measurements from the $Q_{\rm T}(r,\phi)$ map and the Fourier analysis have the strongest correlations in both the bar strength (Figure \ref{fig3.5.1}(b)) and length (Figure \ref{fig3.5.1}(e)). The scatter in the bar strength is larger compared to that in the bar length. We separate galaxies into early-type (S0/a-Sb), intermediate-type (Sbc-Scd), and late-type (Sd-Sm) spirals by red solid circles, green open triangles, and blue solid squares, respectively, in Figure \ref{fig3.5.1}. We find a trend that for a given $Q_{\rm b}$ value early-type spirals are skewed to larger $A_2$ values, whereas late-type spirals to smaller $A_2$ values. These results have been shown in previous studies and explained by the presence of a large bulge in early-type spirals and a dominant dark matter halo in late-type spirals \citep{2002Laurikainen_Salo, 2004Buta, 2004aLaurikainen, 2016Diaz}. We will discuss it further in \S\ref{chap3.5.2}, \S\ref{chap3.6.1}, and \S\ref{chap6.2}. 

On the other hand, the measurements by the ellipse fitting method and Fourier analysis are least-correlated in both the bar strength (Figure \ref{fig3.5.1}(c)) and the bar length (Figure \ref{fig3.5.1}(f)). The number of plotted samples is reduced because we only plot galaxies that satisfy the bar criteria in both methods. 

The bar force ratio $Q_{\rm b}$ also has a good correlation with the bar ellipticity $\epsilon_{\rm bar}$ (Figure \ref{fig3.5.1}(a)), which reflects the fact that the underlying potential basically determines the orbits of stars \citep{2016Diaz}. It supports the proposition that the maximum ellipticity of a bar $\epsilon_{\rm bar}$ can be a good approximation of the bar force ratio $Q_{\rm b}$ \citep{1992Athanassoula, 2002Laurikainen, 2002Laurikainen_Salo}. 

However, the relation of $Q_{\rm b}$ and $\epsilon_{\rm bar}$ looks different from those between other parameters that are roughly linear (Figures \ref{fig3.5.1}(b) and (c)). It is because the bar force ratio $Q_{\rm b}$ is related to the bar mass fraction and the orbit of stars \citep{2012Kim}. \citet{2012Kim} analytically derived a simple relation between $Q_{\rm b}$, the bar mass fraction $f_{\rm b}$, and the axis ratio $a/b$, where $a$ and $b$ are the semi major axis and minor axis, for a prolate Ferrers bar assuming a flat rotation curve. The force ratio $Q_{\rm b}$ turned out to be a nonlinear, yet simple, function of the bar mass fraction $f_{\rm b}$ for a given degree of central density concentration $c$:  
\begin{eqnarray}
Q_{\rm b} &=& 0.58 f_{\rm b}^{0.89}(a/b - 1) \\
          &=& 0.58 f_{\rm b}^{0.89}\frac{\epsilon_{\rm bar}}{1-\epsilon_{\rm bar}},   \qquad{\rm for} \; c = 0
\end{eqnarray}
or
\begin{eqnarray}
Q_{\rm b} &=& 0.44 f_{\rm b}^{0.87}(a/b - 1) \\
          &=& 0.44 f_{\rm b}^{0.87}\frac{\epsilon_{\rm bar}}{1-\epsilon_{\rm bar}},   \qquad{\rm for} \; c = 1
\end{eqnarray}
\citep{2012Kim}. The bar mass fraction $f_{\rm b}$ is defined as $M_{\rm bar}/(M_{\rm bar}+M_{\rm bulge})$ where $M_{\rm bulge}$ is measured by the bulge mass inside $r = 10$ kpc. The central density concentration $c = 0$ denotes a homogeneous bar, whereas $c = 1$ means a more concentrated bar. Comparing with the observational results of \citet{2010Comeron}, \citet{2012Kim} suggested that bars in real galaxies are unlikely to be more concentrated than $c = 1$ and might have a bar mass fraction of $0.25$ to $0.5$. We also compare these analytic relations with our data: they are well represented by a homogeneous bar ($c = 0$) with $f_{\rm b} = 0.4$ to $0.5$. We overplot the relation with $f_{\rm b} = 0.4$ (gray line) and $f_{\rm b} = 0.5$ (black line) for $c = 0$ in Figure \ref{fig3.5.1}(a). 

When it comes to the bar length measurements, $r_{\rm Qb}$ shows the strongest correlation with $r_{\rm A2}$ (Figure \ref{fig3.5.1}(e)). The $Q_{\rm T}(r,\phi)$ map and Fourier analysis are quite consistent with each other in measuring the bar length, as reported in the previous study \citep{2004aLaurikainen}. We do not find any skewed distribution between early- and late-type spirals in the bar length measurements (Figure \ref{fig3.5.1}(e)), differently shown in the bar strength measurements (Figure \ref{fig3.5.1}(b)). The outliers often come from confusions by the Fourier analysis misidentifying elongated bulges as bars. On the other hand, we find that even very long bars often show the maximum $Q_{\rm b}$ at the very inner regions of the bar. 

In Figures \ref{fig3.5.1}(d) and (f), we find that the correlation between $r_{\epsilon}$ against $r_{\rm Qb}$ or $r_{\rm A2}$ is not as tight as the correlation between $r_{\rm Qb}$ and $r_{\rm A2}$. According to \citet{2016Diaz}, $r_{\epsilon}$ best matches the bar length estimated by visual inspection. In our analysis, $r_{\rm Qb}$ and $r_{\rm A2}$ are usually measured to be shorter by 20\% than $r_{\epsilon}$. 

\subsection{Dependence of Bar Properties on Galaxy Properties} \label{chap3.5.2}

\begin{figure*}[htbp]
\includegraphics[bb = 10 460 575 790, width = 0.99\linewidth, clip=]{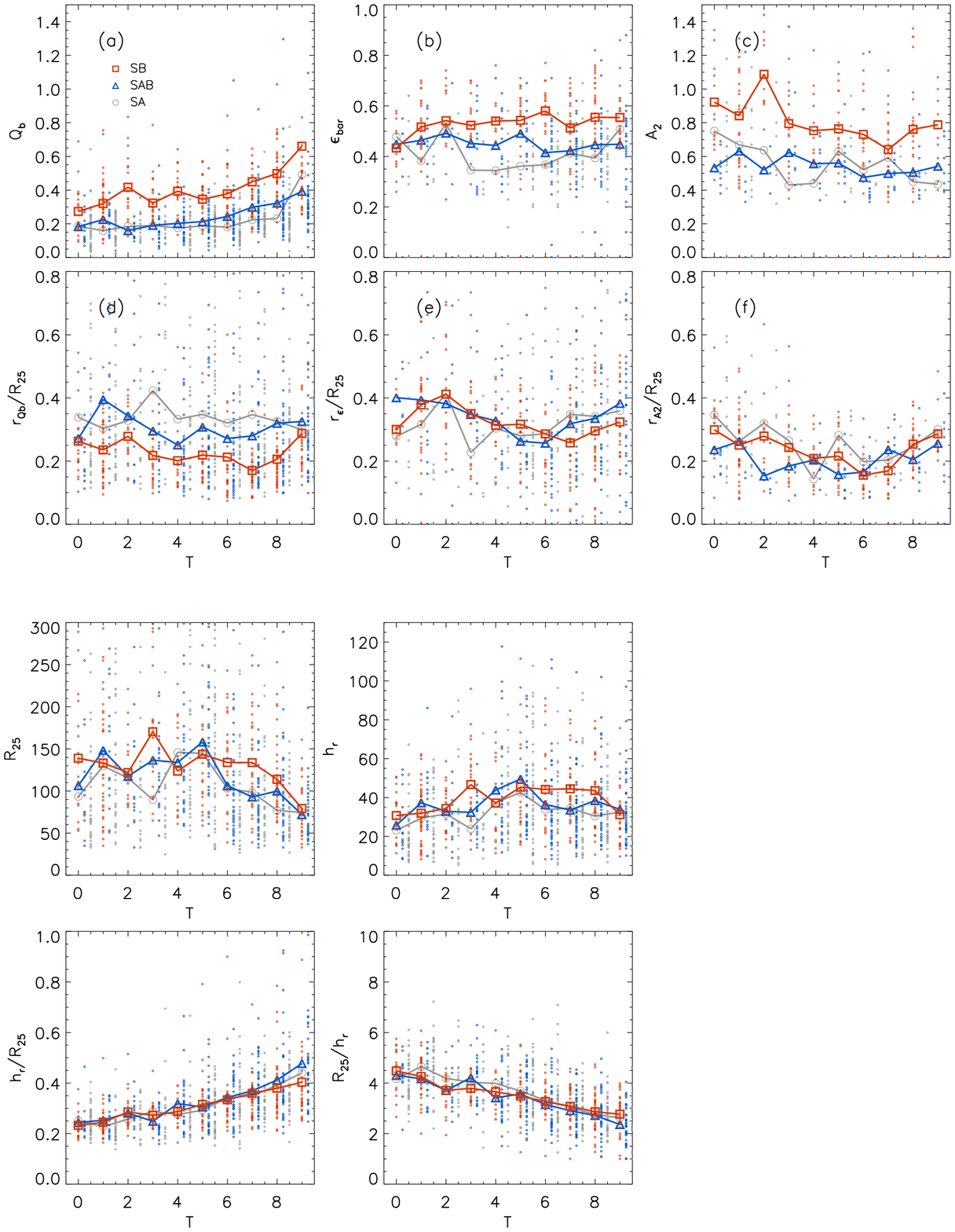}
\caption{The bar strengths ({\it upper row}) and bar lengths normalized by $R_{25}$ ({\it bottom row}) as a function of the Hubble sequence. Three different measurements, $Q_{\rm b}$ ({\it left}), $\epsilon_{\rm bar}$ ({\it middle}), and $A_2$ ({\it right}), are used. The visually determined SBs, SABs, and SAs are displayed by red, blue, and gray colors with $+0.25$ offsets in the abscissa. The lines show mean values. \label{fig3.5.2}}
\end{figure*}

In Figure \ref{fig3.5.2}, we plot the bar strength and length as a function of the coded numerical revised Hubble stage $T$. We again used the three methods of measurement, the $Q_{\rm T}(r,\phi)$ map analysis, the ellipse fitting method, and the Fourier analysis. In each panel, we distinguish visually determined SBs, SABs, and SAs by \citetalias{Ann15} with successive offsets of $+0.25$ in the abscissa as red, blue, and gray dots, respectively. The lines indicate the mean values at the original bin. With $Q_{\rm T}(r,\phi)$ map analysis, we can measure the strength and length for all galaxies, whether or not the galaxy has a bar, as explained in \S\ref{chap3.3.1}. However, in other methods, we can estimate them only when the methods can detect bars. We note that the Fourier analysis rarely detects SABs \citepalias{2019Lee}

We show the bar strength versus the Hubble type in the upper row of Figure \ref{fig3.5.2}. The individual values are quite scattered, but the mean values show some dependence on the Hubble type. Even though the three bar strength measurements show some degree of correlation to each other (Figure \ref{fig3.5.1}), they do not show consistent tendencies as a function of the Hubble sequence. The mean values of $Q_{\rm b}$ increase toward the late-type spirals with high $T$ (Figure \ref{fig3.5.2}(a)), whereas the mean values of $A_2$ show the opposite trend, decreasing toward the late-type spirals (Figure \ref{fig3.5.2}(c)). The mean values of $\epsilon_{\rm bar}$ show no systematic dependence on the Hubble type (Figure \ref{fig3.5.2}(b)). The Hubble type dependence in $Q_{\rm b}$, increasing toward late-type spirals, has been previously reported \citep{2002Laurikainen, 2002Whyte, 2004Buta, 2010bButa} and even the opposite tendencies between $Q_{\rm b}$ and $A_2$ also have been studied \citep{2004aLaurikainen, 2016Diaz}. 

SABs have lower bar strengths compared to SBs. The value of $Q_{\rm b}$ separates SBs from SABs or SAs most consistently compared to $A_2$ or $\epsilon_{\rm bar}$. The bar ellipticity $\epsilon_{\rm bar}$ is worst in separating SBs from SABs or SAs. However, the opposite tendencies for T do not differ in SABs and even in SAs. 

In fact, we can expect these opposite distributions from Figure \ref{fig3.5.1}(b) where early-type spirals are skewed to larger $A_2$ and late-type spirals to larger $Q_{\rm b}$. This can be most likely explained by the bulge dilution: the underlying radial force generated by a bulge causes underestimation of $Q_{\rm b}$, especially in early-type spirals \citep{2001Block, 2002Laurikainen_Salo, 2004aLaurikainen, 2004Buta, 2007Elmegreen, 2016Diaz}. We test how much a bulge affects the estimation of the bar strength depending on the methods by using mock galaxies with different bulge-to-total ratios in \S\ref{chap3.6.1}.

Next, we display the relative bar length with respect to $R_{25}$ at the bottom row of Figure \ref{fig3.5.2}. The previous study shows that early-type spirals have longer bars in both absolute and relative lengths with respect to $R_{\rm 25}$ than late-type spirals \citep{1985Elmegreen, 1995Martin, 2005Erwin, 2007Elmegreen, 2007Menendez, 2009Aguerri}. However, most of them used a small sample of galaxies or did not use the whole Hubble types. We find that early-type spirals have long bars, but the relative bar length increases from intermediate-type to late-type spirals as well. This result is consistent with the previous studies that considered the whole Hubble types, which show a bimodal distribution that the relative bar length increases towards both ends of the Hubble sequence with a minimum at T = 5 $\sim$ 6 \citep{2007Laurikainen, 2016Diaz, 2017Font, 2019Font}. It means that the bar length does not decrease as much as the disk size decreases in late-type spirals. This tendency looks similar for all three different measurements, and both for SBs and SABs. 

Lastly, we compare the bar length between SBs and SABs. Despite the general understanding that bars in SBs are highly contrasted, stronger, and longer compared to those in SABs \citep{1985Elmegreen, 1987Ann, 1989Elmegreen, 1990Ohta, 2005Erwin, 2009Laurikainen, 2012Lee}, we can not find in our analyses the evidence that bars in SBs are longer than those in SABs. The mean values of the relative bar length of SBs are not definitively greater than those of SABs in the same bin of Hubble type. In terms of $r_{\rm Qb}$, the mean bar lengths of SABs are even larger than those of SBs (Figure \ref{fig3.5.2}(d)). Accordingly, when we compare the results for $Q_{\rm b}$ and $r_{\rm Qb}/R_{25}$ in Figures \ref{fig3.5.2}(a) and (d), it is likely that the maximum point of the bar force ratio $r_{\rm Qb}$ is moving inwards as the bar is growing stronger. We find similar behaviour that $r_{\rm Qb}$ decreases as $Q_{\rm b}$ increases in a previous study \citep[see Figure 5]{2002Laurikainen} and also in the comparison with the simulation data by \citet{2019Seo}, which is shown in Figure \ref{fig3.6.2} of \S\ref{chap3.6.2}.

\section{Discussion} \label{chap3.6}

\subsection{Bulge Effect on Bar Properties} \label{chap3.6.1}

Bars in early-type spirals have been believed to be longer and stronger than those in late-type spirals \citep{1985Elmegreen, 1987Ann, 1989Elmegreen, 1990Ohta, 1996Elmegreen, 2004aLaurikainen, 2005Erwin, 2000Abraham, 2009Laurikainen, 2012Lee}. Indeed, in terms of the Fourier amplitude, early-type spirals have stronger bars than late-type spirals (Figure \ref{fig3.5.2}(c)), which is consistent with previous studies \citep{1990Ohta, 2004aLaurikainen, 2009Laurikainen}. However, in terms of the bar force ratio and bar ellipticity, it is hard to say that bars in early-type spirals are stronger than those in late-type spirals (Figures \ref{fig3.5.2}(a) and (b)). This has also been reported in the literature \citep{2002Laurikainen, 2004Buta, 2010Buta, 2004aLaurikainen, 2016Diaz}.  

In order to understand the lower bar force ratio in early-type spirals, \citet{2004Buta} investigated several possible causes. They first took into account more spherical shapes of bulges by using two-dimensional decompositions for bulges, disks, and bars. The assumption for the bulge to be the same thin shape as the disk overestimates the radial force in the plane \citep{2002Laurikainen_Salo, 2004Buta}. Secondly, they applied a constant ratio of $h_r/h_z = 12$ for all Hubble type galaxies instead of the type dependent $h_r/h_z$ because high $h_r/h_z$ in late-type spirals enhances $Q_{\rm b}$. Thirdly, they tried to separate $Q_{\rm b}$ from the force ratio by spiral arms. Lastly, \citet{2004Buta} and \citet{2016Diaz} found that the dark halo correction lowers $Q_{\rm b}$ by $20 \sim 25\%$ in T = $7 - 10$. However, the dark halo could not fully explain the Hubble type dependence of $Q_{\rm b}$. The $Q_{\rm b}$ has still turned out to be lower in early-type spirals than in late-type spirals. We will discuss this more in \S\ref{chap6.2}.

\begin{figure*}[htbp]
\includegraphics[bb = 10 455 580 790, width = 0.99\linewidth, clip=]{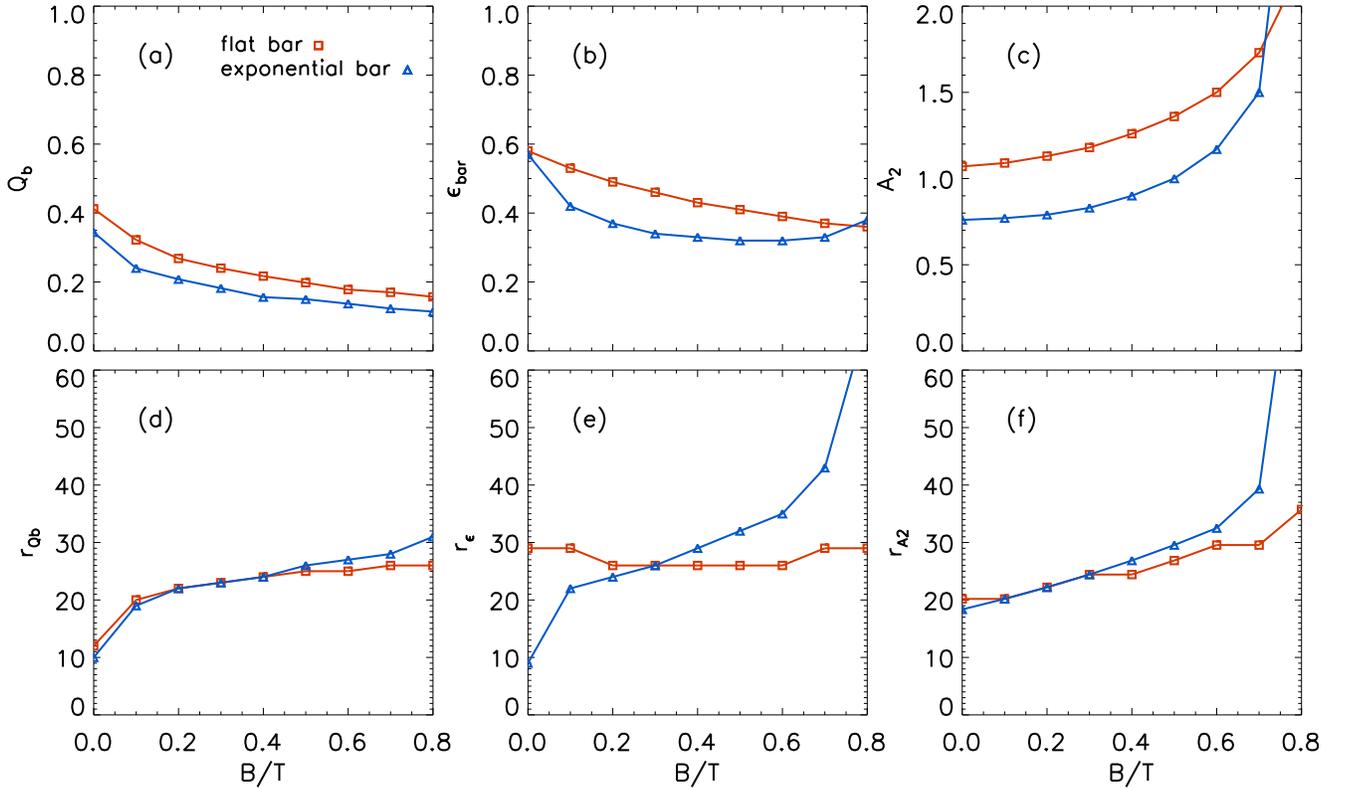}
\caption{The bar strengths ({\it upper row}) and lengths ({\it bottom row}) as a function of the bulge-to-total ratio for mock galaxies with a bulge, a disk, and a bar. The bar strength and length are estimated by three different methods: $Q_{\rm b}$ and $r_{\rm Qb}$ from $Q_{\rm T}(r,\phi)$ map analysis ({\it left}), $\epsilon_{\rm bar}$ and $r_{\epsilon}$ from ellipse fitting method ({\it middle}), and $A_2$ and $r_{A_2}$ from Fourier analysis ({\it right}). Red squares and blue triangles indicate flat barred galaxies and exponential barred galaxies, respectively. \label{fig3.6.1a}}
\end{figure*}

On the other hand, it has been suggested that a bulge significantly dilutes $Q_{\rm b}$ in early-type spirals, for it is an important contributor to the radial force in the bar region \citep{2001Block, 2002Laurikainen_Salo, 2004Buta, 2004aLaurikainen, 2007Elmegreen, 2016Diaz}. \citet{2004aLaurikainen} reported by investigating 180 disk galaxies that $Q_{\rm b}$ varies from 0.6 to 0.1 as the bulge-to-total ratio increases from 0 to 1. 

Here, we test the bulge effect on the value of the bar strength measured by all three different methods. We constructed various mock galaxies with the bulge-to-total ratio (B/T) of 0 to 0.8. A bulge, a disk, and a bar with the flat or exponential profile were built as described in \S\ref{chap3.3.2}. In Figures \ref{fig3.6.1a}(a)-(c), we display each measurement of the bar strength $Q_{\rm b}$, $\epsilon_{\rm bar}$, and $A_2$ as a function of B/T for these mock galaxies. Red squares and blue triangles indicate flat and exponential barred galaxies, respectively. We find that not only $Q_{\rm b}$ but also $\epsilon_{\rm bar}$ and $A_2$ largely depend on B/T. When B/T increases from 0 to 0.8, $Q_{\rm b}$ drops to less than half and $\epsilon_{\rm bar}$ reduces by around $30\%$ (Figures \ref{fig3.6.1a}(a)-(b)). In contrast, $A_2$ increases over by $50\%$ (Figure \ref{fig3.6.1a}(c)). This experiment explains the different Hubble type dependence of bar strength measured by different methods shown in Figure \ref{fig3.5.2}, confirming the bulge dilution on $Q_{\rm b}$ suggested by the previous studies \citep{2001Block, 2002Laurikainen_Salo, 2004aLaurikainen, 2007Elmegreen, 2016Diaz}. 

We emphasize that the bulge effect is as significant as the effect of the bar. It means that the measured bar strength for the same bar may decrease to half the value depending on the bulge that resides with the bar in the host galaxy. Besides, depending on the method to estimate the bar strength, the same bulge can decrease the bar strength $Q_{\rm b}$ and $\epsilon_{\rm bar}$ or increase the bar strength $A_2$.

\begin{figure*}[htbp]
\includegraphics[bb = 20 455 580 790, width = 0.9\linewidth, clip=]{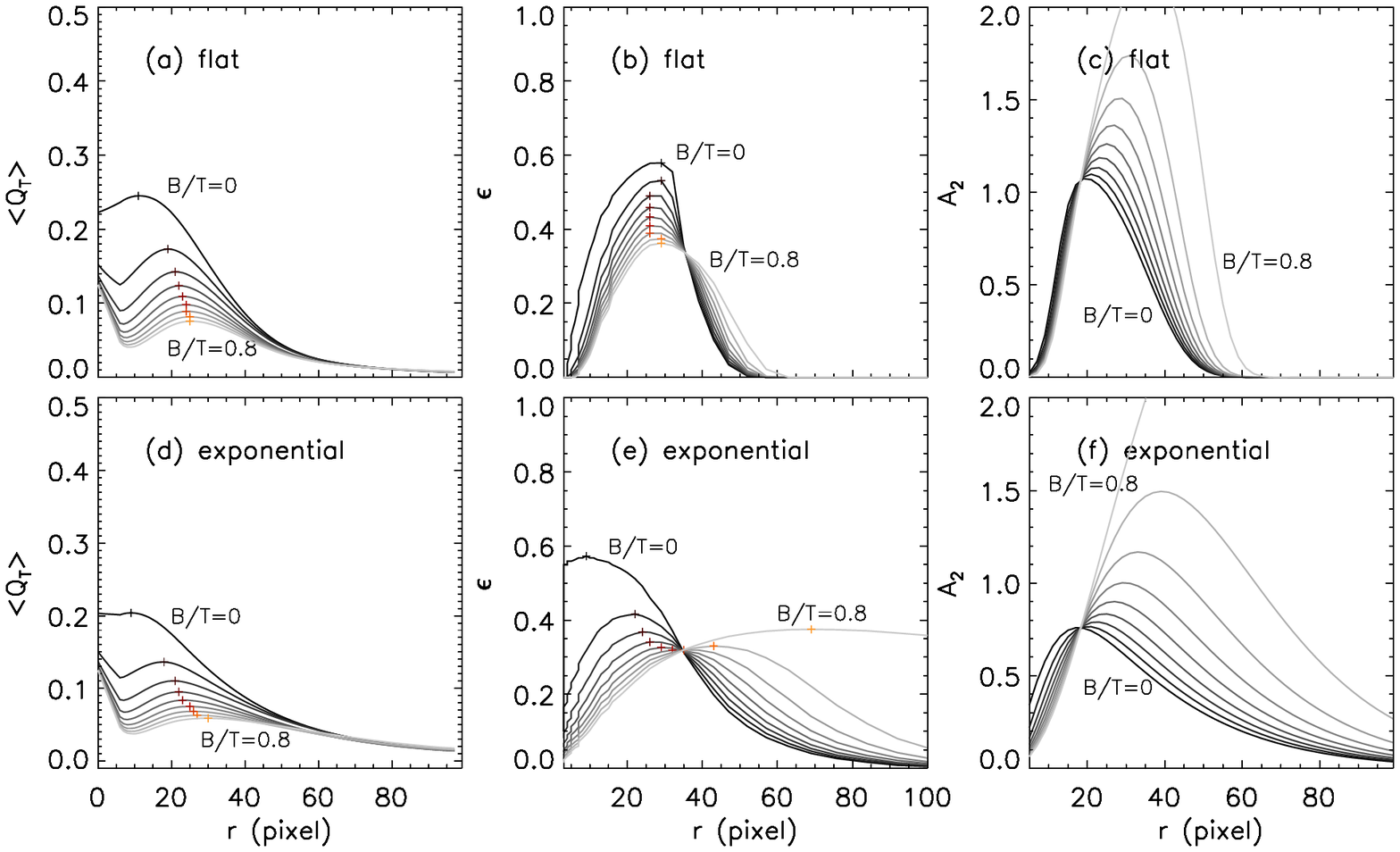}
\caption{Radial profiles of $\langle Q_{\rm T} \rangle$ ({\it left}), $\epsilon$ ({\it middle}), and $A_2$ ({\it right}) for all individual mock galaxies. Lighter colors indicate larger B/T, from 0 to 0.8. We plot the flat barred galaxies in the upper row and exponential galaxies in the bottom row. The cross symbols show the location of the maximum bar strength for each mock galaxy. \label{fig3.6.1b}}
\end{figure*}

In Figure \ref{fig3.6.1b}, we plot the radial profiles of $\langle Q_{\rm T} \rangle$, $\epsilon$, and $A_2$ for all mock galaxies. The upper row shows flat barred galaxies, and the bottom row exponential barred galaxies. We indicate the B/T ratio by the colors of lines: as the colors become lighter, B/T increases from 0 to 0.8. The cross symbols show the loci of the maximum bar strength. We confirm that when the color becomes lighter (B/T increases), $Q_{\rm b}$ and $\epsilon_{\rm bar}$ decrease, while $A_2$ increases. When it comes to the bar length, the locations of the maximum strength tend to move outward in most cases. These tendencies similarly appear in both flat and exponential barred galaxies. 

\begin{figure}[htbp]
\centering
\includegraphics[bb = 10 285 400 635, width = 0.99\linewidth, clip=]{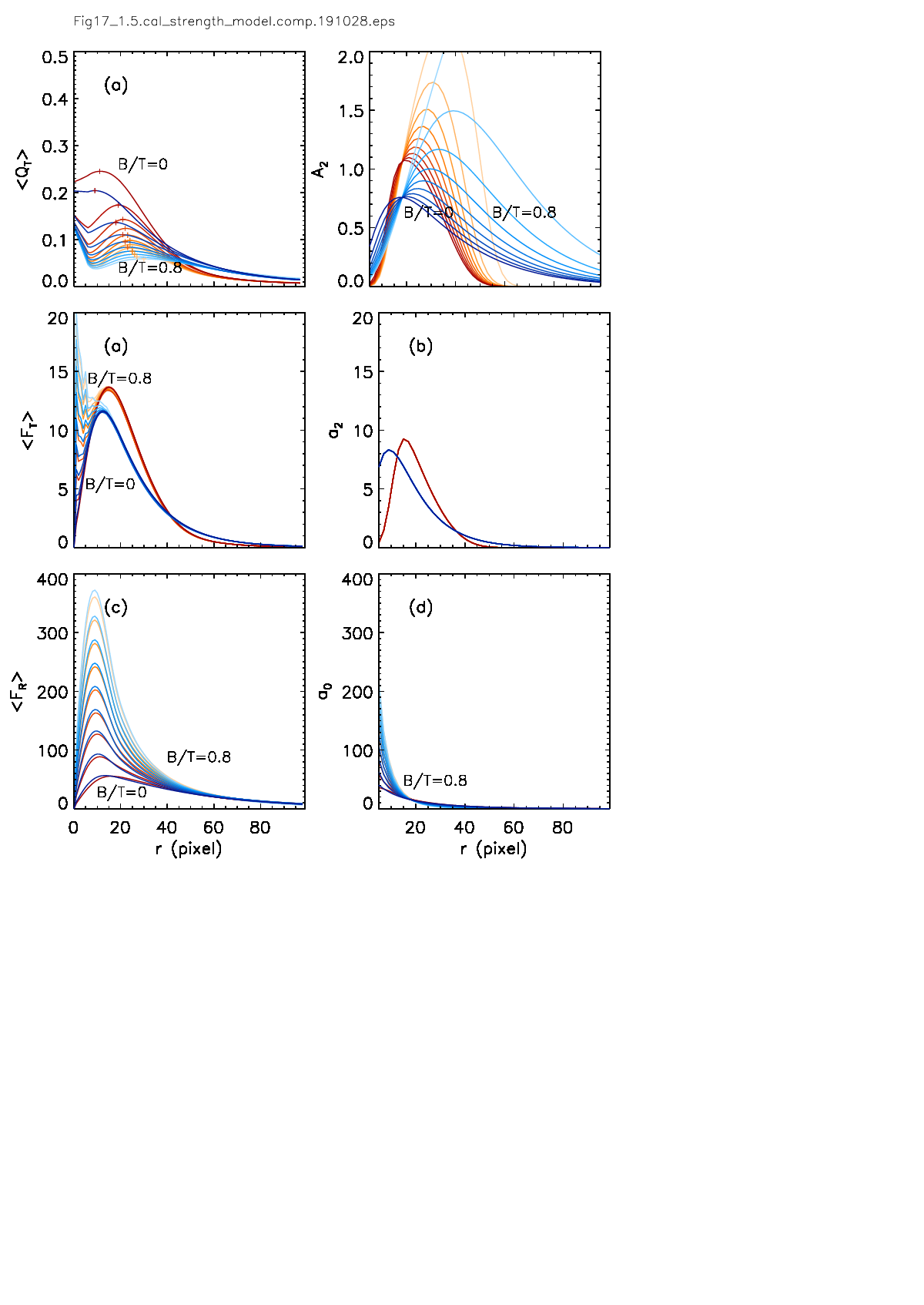}
\caption{The components of $\langle Q_{\rm T} \rangle$ ({\it left}) and $A_2$ ({\it right}) for all individual mock galaxies. The upper row shows the radial profiles of $\langle F_{\rm T} \rangle$ and $a_2$, and the bottom row $\langle F_{\rm R} \rangle$ and $a_0$. Red and blue colors mean the components in flat and exponential barred galaxies. As colors become lighter, B/T increases from 0 to 0.8. \label{fig3.6.1c}}
\end{figure}

To analyze the cause of the opposite trend between $Q_{\rm b}$ and $A_2$, we decomposed the profiles of $\langle Q_{\rm T} \rangle$ and $A_2$ into the components, $\langle F_{\rm T} \rangle$, $\langle F_{\rm R} \rangle$, $a_2$, and $a_0$ in Figure \ref{fig3.6.1c}. We indicate the components in flat and exponential bars by red and blue colors, respectively. Lighter colors mean larger B/T as well. In Figures \ref{fig3.6.1c}(a) and (b), we confirm that the profiles of tangential force $\langle F_{\rm T} \rangle$ and $a_2$ are hardly affected by the bulge ratio. They are the representative components of a bar. What we have to be cautious is that the radial force $\langle F_{\rm R} \rangle$ and $a_0$ are more strongly affected by the bulge than we have expected (Figures \ref{fig3.6.1c}(c) and (d)). The higher B/T totally changes the profiles of $\langle F_{\rm R} \rangle$ and $a_0$, which results in the different bar strength and location of the maximum strength, i. e., bar length. We find that the opposite trends between $Q_{\rm b}$ and $A_2$ have been caused by the changing profiles of $\langle F_{\rm R} \rangle$ and $a_0$ as B/T changes. 

Therefore, we need to consider the bulge effect on the bar strength and length more carefully. Fortunately, in the case of the bar length, the bulge effect seems less significant compared to the bar strength (Figures \ref{fig3.6.1a}(d)-(f)). However, we should note that when the B/T increases, the bar force ratio $Q_{\rm b}$ decreases but the bar length $r_{\rm Qb}$ becomes longer. 

When we compare the types of bars, the exponential bars are more significantly affected by the bulge compared to the flat bars (Figures \ref{fig3.6.1a} and \ref{fig3.6.1b}). In particular, for the ellipse fitting, larger bulges in flat barred galaxies make the shape of a bar rounder and lower the ellipticity in the bar region (Figure \ref{fig3.6.1b}(b)). Meanwhile, in exponential barred galaxies, ellipses are mainly determined by a bulge rather than a bar (Figure \ref{fig3.6.1b}(e)). When the B/T increases, $\epsilon_{\rm bar}$ decreases and $r_\epsilon$ is largely pushed outward.  

\subsection{Variable Mass-to-light Ratio on Bar Force Ratio}\label{chap6.2}

We have calculated the gravitational potential by assuming constant mass-to-light ratio (M/L) over the disk \citep{1994Quillen}. The stellar M/L is a vital parameter to translate photometry to dynamics \citep{2001Bell}, and inevitably affected by the dark matter. We used $i-$band images to be less affected by the stellar population, but there are still variations in stellar M/L within and among galaxies as large as a factor of $3$ in $I$ band \citep{2001Bell}. \citet{2001Bell} showed that the stellar M/L has a strong correlation with the color of the integrated stellar population, and \citet{2004Buta} have suggested a way to consider the color-dependent M/L in estimating the potential because of the radial stellar population variations, in particular, in late-type spirals. However, they did not choose to use the approach because of the impractical reason to adopt it to the full sample.

Here, we examine the dark matter effect on $Q_{\rm b}$ by adopting the color-dependent M/L instead of the constant M/L. We have inferred the effective stellar mass at each pixel in $i-$band images with $(g-i)$ color maps, which gives the best agreement with those using optical/NIR color such as $(i-H)$ \citep{2009Zibetti}. We have constructed $(g-i)$ color maps for 120 galaxies, excluding objects that needed special reductions, among visually determined strongly barred galaxies in our sample. We adopt the relation between stellar M/L and color $(g-i)$, ${\rm log}_{\rm 10}(M/L) = -0.152+0.518 \times (g-i)$ \citep{2003Bell}. 

\begin{figure}[htbp]
\includegraphics[bb = 30 590 300 800, width = 0.95\linewidth, clip=]{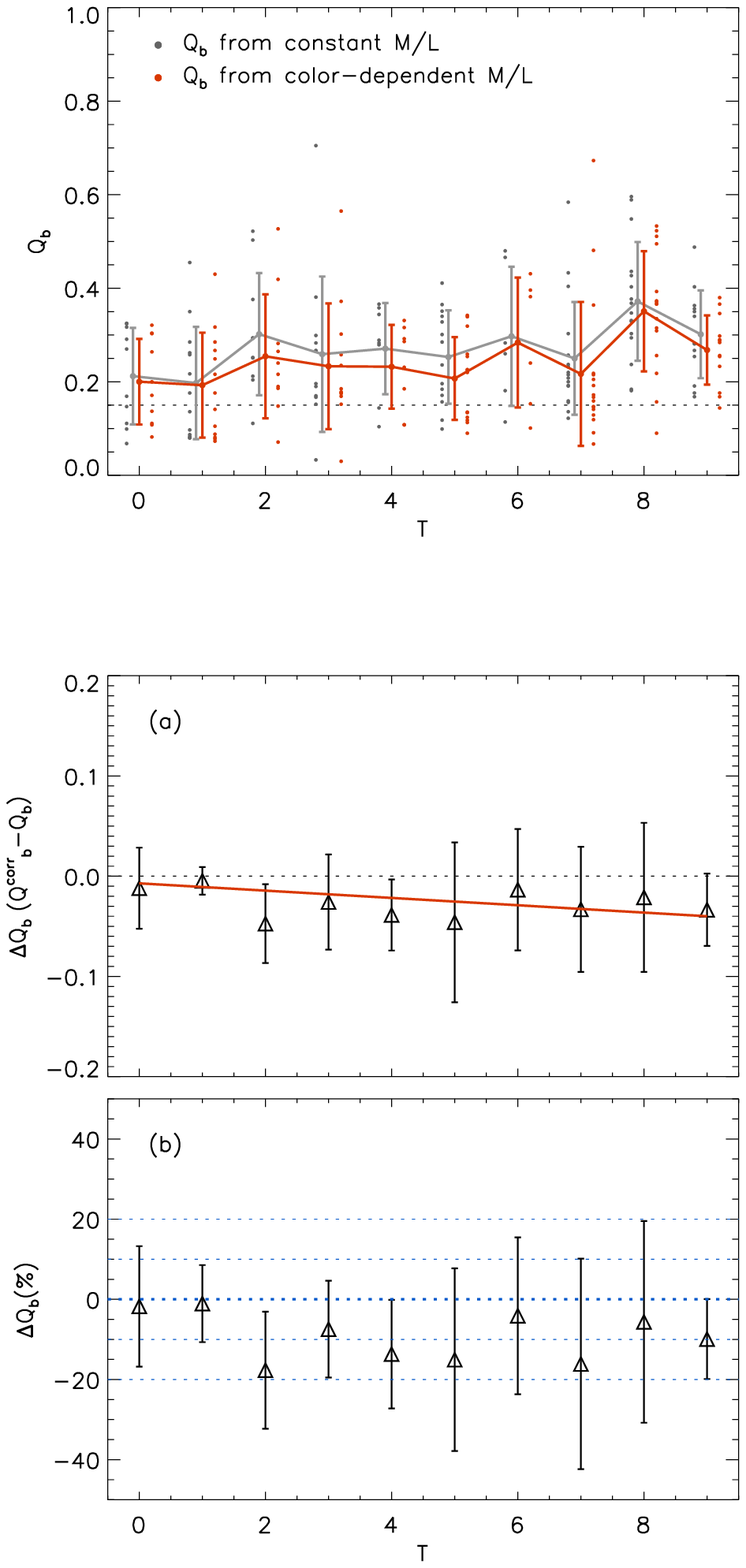}
\caption{The force ratio $Q_{\rm b}$ as a function of the Hubble sequence. The red circles indicate $Q_{\rm b}$ obtained from adopting the constant M/L, and blue triangles mark corrected $Q_{\rm b}$ by applying the color-dependent M/L. The lines show the mean values. \label{fig6.2a}} 
\end{figure}

In Figure \ref{fig6.2a}, we display the force ratio from constant M/L (gray circle) against color-dependent M/L (red circle) as a function of the Hubble sequence. The mean $Q_{\rm b}$ from constant M/L increases toward late-type spirals, although the values are lower than that for the whole sample as shown in Figure \ref{fig3.5.2}(a). The color-dependent M/L tends to lower the mean $Q_{\rm b}$ over the whole Hubble sequence. As expected, the mean difference is minimal in T = 0 and T = 1, and becomes large, but within 20\%, in other Hubble types. The difference increases by $0.004$ on average in the direction of lower $Q_{\rm b}$ as T increases by one. However, it does not change the general tendency of increasing $Q_{\rm b}$ toward late-type spirals. It is consistent with previous studies by \citet{2004Buta} and \citet{2016Diaz} who examined the dark matter effect on $Q_{\rm b}$ using the universal rotation curve models. Compared to the results by \citet{2016Diaz}, the color-dependent M/L reflects the characteristics of individual galaxies such as the stellar population and dust lanes. \citet{2016Diaz} showed more systematic reduction of the mean $Q_{\rm b}$ by about 10-15\% in $3 \le T < 5$ and 20-25\% in T = 7-10. 

\subsection{Comparison with a Numerical Simulation} \label{chap3.6.2}

In \S\ref{chap3.4.1}, we introduced two types of the radial profile $\langle Q_{\rm T}\rangle(r)$: one that has a maximum peak and the other a plateau on the radial profile (Figures \ref{fig3.4.1a} and \ref{fig3.4.1b}). We label them type M or P, respectively. We have classified our 884 sample galaxies into type M and P. Sometimes, a maximum peak looks like a plateau because the difference between the maximum and the minimum is minimal. However, as long as the profile has a maximum peak, we classified them as type M. In Table \ref{Table3.6.2}, we organize the fractions of type M and P for barred and nonbarred galaxies. We found the same fractions for type M (50\%) and P (50\%) from the whole sample. For barred galaxies, type M (33\%) is more common than type P (20\%), while, in nonbarred galaxies, type P (30\%) is more frequent than type M (17\%).  

\begin{table}[htb]
 \caption{The Fractions of Type M and Type P Galaxy \label{Table3.6.2}}
\begin{center}
\begin{tabular}{l|cc|c}
\hline
 {bar/nonbar} &
 {type M} &
 {type P} &
 {total} \\ 
\hline
bar    & 294 (33\%) & 174 (20\%) & 468 (53\%) \\
nonbar & 148 (17\%) & 268 (30\%) & 416 (47\%) \\ 
\hline
total  & 442 (50\%) & 442 (50\%) & 884 (100\%) \\ 
\hline
\end{tabular}
\end{center}
\end{table}

We were curious why the radial profiles appear to have two different types. In fact, we did not find any type P galaxies among our mock galaxies that have a flat or an exponential bar in Figures \ref{fig3.6.1b}(a) and (d). Our visual inspection on the sample galaxies revealed that type M galaxies tend to have more developed bars along with other structures such as grand-design spiral arms or rings. On the other hand, type P galaxies rarely show prominent structures, except for bars. Thus, we suspect that the features in the radial profile might be related to the stage of galaxy evolution.

We have applied the $Q_{\rm T}(r,\phi)$ map analysis to the numerical simulation from \citet{2019Seo}. They ran self-consistent simulations of Milky Way-sized, isolated disk galaxies to explore the effect of a cold and warm disk in bar formation and evolution. The initial galaxy models are comprised of a stellar disk, a gaseous disk, a dark matter halo, and a central supermassive black hole. They considered relatively cold and warm disks with different velocity anisotropy parameters $f_R \equiv \sigma ^2 _{R}/\sigma^2_{z} = 1.0$ and $f_R = 1.44$, respectively. The total disk mass is fixed at $M_{\rm disk} = 5 \times 10^{10} M_\odot$, but the gas fraction varies from $0\%$ to $10\%$. The masses of a dark matter component and a supermassive black hole are $M_{\rm DM} = 3.1 \times 10^{11}M_\odot$ and $M_{\rm BH} = 4 \times 10^6 M_\odot$, respectively.

\begin{figure*}[htbp]
\includegraphics[bb = 20 80 530 780, width = 0.9\linewidth, clip = ]{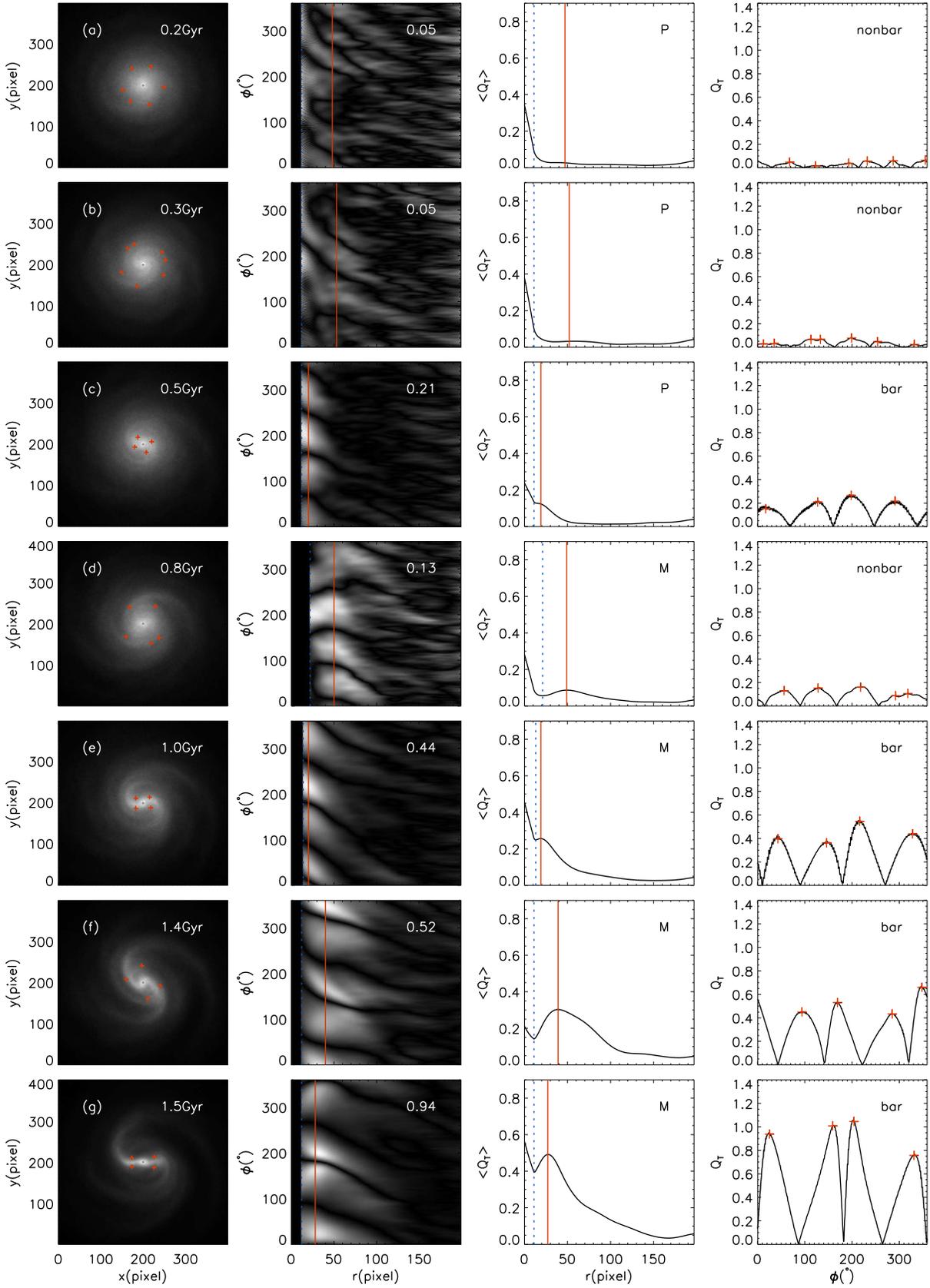}
\caption{The $Q_{\rm T}(r,\phi)$ map analysis for the simulated galaxies that has a warm disk with 5\% gas fraction from \citet{2019Seo}. They are snapshots between 0 Gyr and 5 Gyr from top to bottom. Each snapshot is analyzed by the $Q_{\rm T}(r,\phi)$ map, radial profile, and azimuthal profile from left to right. We present the time scales, bar strengths $Q_{\rm b}$, M or P type, and bar or nonbar at the top right from left to right. \label{fig3.6.2}}
\end{figure*}

Figure \ref{fig3.6.2} shows the $Q_{\rm T}(r,\phi)$ map analysis for the warm-disk model with 5\% gas fraction, which is considered to be the closest model to the Milky Way \citep{2019Seo}. The first column shows the stellar surface density in logarithmic scale in the 10 kpc region. They are snapshots of the entire simulation from 0 Gyr to 5 Gyr. The simulation showed that the warm disk starts to form a weak bar at $t$ = 0.4 Gyr, and the bar grows stronger until $t$ = 2.0 Gyr. The simulation often shows a spiral structure only without a bar for a short time at $t$ = 0.8-0.9 and at $t$ = 1.4 Gyr (Figures \ref{fig3.6.2}(d) and (f)) during the bar evolution (At 1.4 Gyr, our $Q_{\rm T}(r,\phi)$ analysis could not distinguish an elongated spiral arm from a bar). The second column presents $Q_{\rm T}(r,\phi)$ maps with the bar strength $Q_{\rm b}$ on the top right. The four thick slabs become longer, and the bar strengths increase until 2 Gyr. After 2 Gyr, the bar becomes a little weak up to 3 Gyr and grows again until 5 Gyr, the end of this simulation \citep{2019Seo}.

In the third column, we observe that the radial profile of $\langle Q_{\rm T}\rangle$ also grows with time. Until 0.3 Gyr (Figures \ref{fig3.6.2}(a) and (b)), the profile does not have any prominent structure except for the most central region. From 0.4 Gyr to 0.7 Gyr, we found a plateau outside the central region (Figure \ref{fig3.6.2}(c)). We classify the galaxy at this stage as type P. At this stage, we find weak bars in the stellar surface density images. The profile $\langle Q_{\rm T} \rangle$ grows to have a maximum peak from 0.8 Gyr to the end of the simulation (Figures \ref{fig3.6.2}(d)-(g)). These stages are categorized as type M. In the stellar surface density images, we confirm that weak bars grow to strong bars, passing through a short stage of the nonbarred galaxy. 

In this simulation, we observe that a disk grows roughly in sequence from a type P nonbar, a type P bar, a type M nonbar, to a type M bar. We may understand that the different features on the radial profiles might be related to the evolutionary stage of a bar. We speculate that a type P bar might be a young bar in the early stage, and a type M bar a more developed bar in the late stage of the bar evolution. This is consistent with our first expectation when we investigated the type M and P galaxies by a visual inspection. In addition, we can surmise that type P galaxies might be related to weakly barred galaxies and type M galaxies to strongly barred galaxies. We will discuss this more in \S\ref{chap3.6.3}. 

In the fourth column, we display the azimuthal profile of $Q_{\rm T}$ at $r_{\rm Qb}$. The peaks grow in their heights with time, and the shapes also change. At 0.5 Gyr (Figure \ref{fig3.6.2}(c)), the shapes of four peaks are round, and the four peaks are equally spaced in $\phi$. However, as the bar grows, the shape of peaks becomes sharper, and the four peaks are coupled into two peaks (Figure \ref{fig3.6.2}(g)). In fact, we find similar coupled peaks in real sample galaxies, as shown in Figure \ref{fig3.4.1a} for NGC 4691. This may be a characteristic of a well-developed bar. 

Lastly, we compare the type fraction of sample galaxies with that expected from the simulation data. In the simulation, the galaxy appears as a type M barred galaxy for 84\% of its lifetime, while it stays at the type P for only 14\% of its lifetime. However, we just find 33\% type M barred galaxies in our observed sample and type P galaxies account for half of the whole sample. Therefore, considering the lifetime in the simulation, the fraction of type M barred galaxies in our sample is too small, whereas type M nonbarred galaxies and type P galaxies are too prevalent. Therefore, there seem to be some other factors that make barred galaxies not grow to type M or weaken to type P in real galaxies. We expect that the simulation study can describe real galaxies better if more accurate physical effects such as the effect of the bulge component is considered.  

\subsection{Bar Classification and Hubble Sequence} \label{chap3.6.3}
\begin{figure*}[htbp]
\includegraphics[bb = 20 420 570 760, width = 0.99\linewidth, clip=]{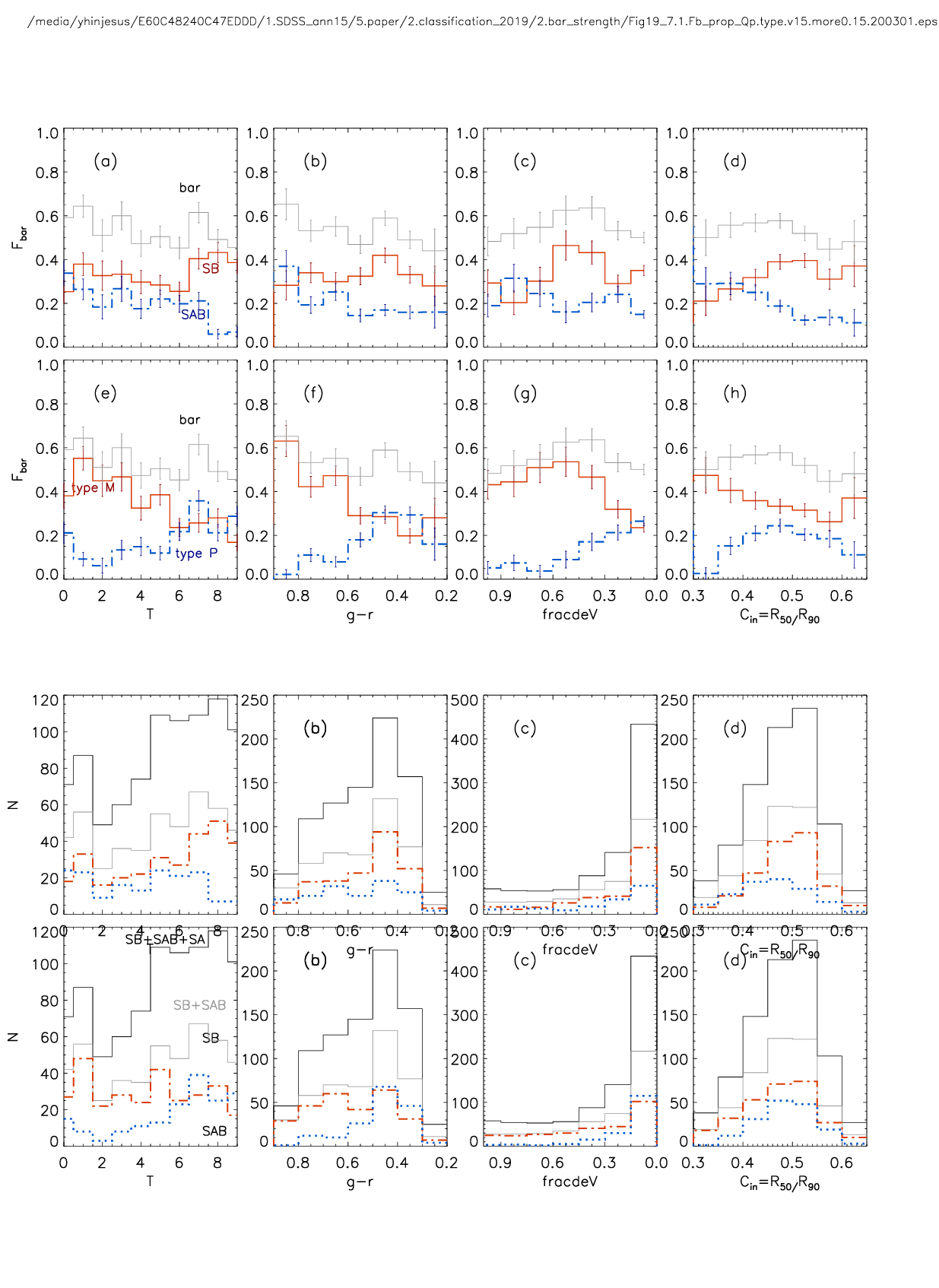}
\caption{Dependence of the fraction of barred galaxies $F_{\rm bar}$ from the $Q_{\rm T}(r,\phi)$ map analysis on the host galaxy properties, the Hubble sequence, $\it g-r$, fracdeV and $C_{in}$. The gray solid lines represent the total barred galaxies by the $Q_{\rm T}(r,\phi)$ map classification. In the top row, SBs (red solid line) and SABs (blue dot-dashed line) are distinguished by the criterion of $Q_{\rm b} = 0.25$, while in the bottom row, type M (red solid line) and type P (blue dot-dashed line) are classified by the features in the radial profile: type M has a maximum peak, whereas type P has a plateau in the profile. We convert the abscissas of $\it g-r$ and fracdeV from right to left in order to arrange characteristics of early-type spirals on the left side in all panels. \label{fig3.6.3}}
\end{figure*}

\citetalias{2019Lee} showed that the bar fraction as a function of the Hubble sequence depends on the method to detect barred galaxies. The bar fraction by the ellipse fitting method increases toward late-type spirals, while that by the Fourier analysis toward early-type spirals. We investigate the bar fractions obtained by our $Q_{\rm T}(r,\phi)$ map analysis as a function of the host galaxy properties such as the Hubble sequence, $\it g-r$, fracdeV, and $C_{\rm in}$ (Figure \ref{fig3.6.3}). The fracdeV indicates the bulge-to-total ratio, which is estimated by the fraction of the light fit by the de Vaucouleurs profile \citep{2010Masters, 2011Masters, 2019Lee}. As the bulge dominates, the fracdeV value increases. The $C_{\rm in}$ denotes the inverse light concentration, $C_{\rm in} \equiv R_{50}/R_{90}$, where $R_{50}$ and $R_{90}$ are the Petrosian radii enclosing 50\% and 90\% of the total galaxy light, respectively \citep{1976Petrosian, 2012Lee}. We used the Hubble types and $\it g-r$ from \citetalias{Ann15}, and other information from the SDSS database. We arrange all parameters in abscissa in the manners that early-type spirals are on the left, and late-type spirals on the right. 

First, we find that the total bar fraction (gray solid lines) does not show any obvious or significant trend as functions of the properties of host galaxies. It looks similar to that of the total bar fraction determined by the visual inspection of \citetalias{Ann15} \citepalias[see Figure 6]{2019Lee}. Previous studies that have dealt with both SBs and SABs show a similar result \citep{1999Knapen, 2000Eskridge, 2017Li}.

On the other hand, in the top row of Figure \ref{fig3.6.3}, when we divide SBs (red solid line) and SABs (blue dot-dashed line) based on the bar force ratio $Q_{\rm b} = 0.25$, we find that SBs generally increase in late-type spirals, but SABs decrease in the same range. This differs from the visual inspection of \citetalias{Ann15}: we reported that SBs are more frequent in early-type spirals, while SABs in intermediate- or late-type spirals in \citetalias{2019Lee}. This is because a large bulge in early-type spirals dilutes the $Q_{\rm b}$ value, as discussed previously. Therefore, early-type barred galaxies are easier to be categorized as SABs compared to late-type barred galaxies. 

Similarly, the bulge effect on the bar strength measurements discussed in \S\ref{chap3.6.1} could also influence the bar fractions as a function of the Hubble sequence differently depending on the bar classification methods, ellipse fitting and Fourier analysis. In Figures \ref{fig3.6.1a}(b) and (c), as B/T increases, the bar ellipticity decreases, whereas the Fourier amplitude increases. It might affect detecting barred galaxies because of the bar criterion $\epsilon_{\rm bar} \ge 0.25$ or $I_2/I_0 \ge 0.3$ \citep{2004Jogee, 2002Laurikainen}. 

In the bottom row of Figure \ref{fig3.6.3}, we investigate the fractions of type M (red solid line) and type P (blue dot-dashed line) galaxies as functions of the host galaxy properties. We suspect that type M barred galaxies are in a more advanced evolutionary state than are type P galaxies, as discussed in \S\ref{chap3.6.2}. The distribution of the fractions of type M and P resembles those of the fractions of SB and SAB by visual inspection, respectively. Type M is more frequent in early-type, red, bulge-dominated, and more concentrated galaxies, while type P is more frequent in late-type, blue, disk-dominated, and less concentrated galaxies. 

\begin{figure}[htbp]
\includegraphics[bb = 30 330 250 510, width = 0.9\linewidth, clip=]{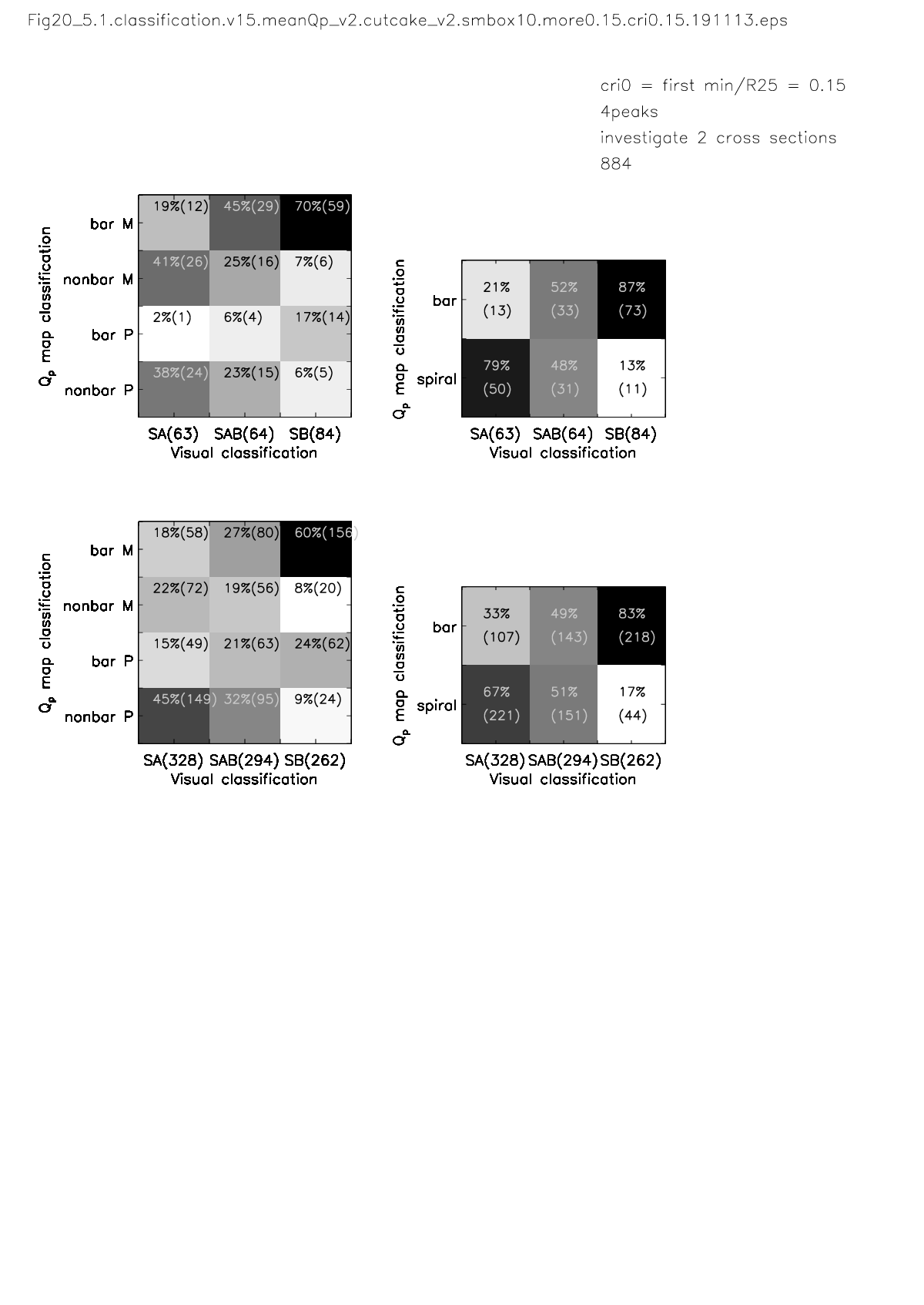}
\caption{Comparison between classifications by analyzing $Q_{\rm T}(r,\phi)$ map and the visual inspection (\citetalias{Ann15}). The $Q_{\rm T}(r,\phi)$ analysis provides the classification into bar or nonbar and type M or P (ordinate). The visual classification is shown in the abscissa. The numbers show how many galaxies classified by the visual inspection are classified into each classification by the $Q_{\rm T}$ map. \label{fig3.6.3b}}
\end{figure}

We examine how the classification of type M and P is related to that of visually classified SB and SAB. In Figure \ref{fig3.6.3b}, we compare our automated classification by $Q_{\rm T}(r,\phi)$ map analysis with the visual classification by \citetalias{Ann15}. In this case, we use the whole sample of galaxies because the concordance sample rarely contains type P galaxies. Our method classifies galaxies into bar or nonbar and type M or P at the same time. The results of the classifications are summarized in Figure \ref{fig3.6.3b}. We confirm that type M barred galaxies are most common in visually classified SB galaxies. Visually determined SAs, on the contrary, have more type P galaxies than type M galaxies. However, we do not find the dominance of visually determined SABs in type P barred galaxies. 

\section{Summary} \label{chap3.7}

We introduced a new method to classify barred galaxies by analyzing the transverse-to-radial force ratio map through several steps: (1) We calculate the radial force and the transverse force by solving the Poisson equation with constant mass-to-light ratio. (2) We transform the ratio map from the Cartesian coordinates $Q_{\rm T}(i,j)$ to the polar coordinates $Q_{\rm T}(r,\phi)$ in order to simplify the butterfly pattern, the bar signature. (3) We investigate the radial profile of the force ratio, averaged over the azimuthal angle, $\langle Q_{\rm T} \rangle (r)$ and find the radius $r_{\rm Qb}$ where the mean force ratio shows a maximum peak or a plateau. (4) We investigate the azimuthal profile $Q_{\rm T}(r_{\rm Qb},\phi)$ at $r_{\rm Qb}$ to see whether it has four peaks, corresponding to the four wings of the butterfly pattern. (5) We automatically classify galaxies as barred galaxies when they have four peaks at the azimuthal profile at $r_{\rm Qb}$ and a bar strength $Q_{\rm b}$ larger than 0.15.  

In the following, we summarize the main results of this work.

\begin{enumerate}
\item This is a first application that utilizes the pattern of the transverse-to-radial force ratio map to classify barred galaxies. We successfully classified barred galaxies of 87\% among visually determined barred galaxies by \citet{RC3} and \citetalias{Ann15} and distinguished nonbarred galaxies with an accuracy of 87\%. We apply this method to 884 spiral galaxies obtained from SDSS/DR7, which are volume-limited with $z < 0.01$ and $M_r < -15.2$, and mildly-inclined with $i < 60^\circ$, selected from 1876 parent sample galaxies. We obtain the bar fraction of 53\%, which is close to the bar fraction ($\sim$60\%) by the classical visual inspection, including both SBs and SABs \citep{1973Nilson, 1987Sandage, RC3, Ann15, 2015Buta}. Our method appears to be the most reliable way to select barred galaxies for large galaxy samples in the era of large galaxy surveys.  

\item Our method provides measurements of the bar strength and the bar length in addition to the bar classification. In particular, we disentangle the bar strength from the spiral strength by analyzing the radial and azimuthal profile of the ratio map. We compare the bar strength $Q_{\rm b}$ and length $r_{\rm Qb}$ from our $Q_{\rm T}(r,\phi)$ map analysis with those from other measurements, the ellipse fitting and Fourier analysis. The bar strength and length measurements show relatively good correlations to each other. Notably, the $Q_{\rm T}(r,\phi)$ map and Fourier analysis are quite consistent with each other in measuring bar strength and length. However, we caution that they yield different tendencies for the Hubble sequence: as the Hubble sequence $T$ increases (later type), $Q_{\rm b}$ increases, whereas the normalized Fourier amplitude $A_2$ decreases. We show the bulge effect is significant in measuring the bar strength and length. In particular, the bulge causes underestimation of the bar strengths $Q_{\rm b}$ and $\epsilon_{\rm bar}$, but overestimation of the bar strength $A_2$ at the same time. When it comes to the bar length, the bulge makes overestimation in all three methods. We point out that we need to consider the bulge effect more carefully when studying the bar strength and the length than previously.  

\item We find that visually determined SBs usually have stronger bar strength $Q_{\rm b}$ than visual SABs. Visual SABs show a similar distribution of $Q_{\rm b}$ with visual SAs rather than with visual SBs. We further subdivided barred galaxies as SBs if $Q_{\rm b} \ge 0.25$ and SABs if $0.15 \le Q_{\rm b} <  0.25$. However, this criterion tends to determine early-type barred galaxies as SABs easily but late-type barred galaxies as SBs because bulges aid to underestimate $Q_{\rm b}$. It shows that the bulge effect on the bar strength measurements can influence the result of the bar classification. Besides, 39\% of visually determined SABs have similar characteristics with nonbarred galaxies in manners that they do not have four peaks on the azimuthal profiles $Q_{\rm T}(r_{\rm Qb}, \phi)$, which supports the suggestion that SAB galaxies are the extension of SA galaxies \citep{2000Abraham, 2019Lee} or SAB galaxies have similar origin of the force ratio with SA galaxies \citep{2004bLaurikainen}. 

\item We find two different types of the radial profile $\langle Q_{\rm T} \rangle (r)$. We classify our sample galaxies that have a maximum peak or a plateau on the radial profiles into type M or P, respectively. We applied the same analysis to numerical simulations of bar evolution by \citet{2019Seo}. The simulations show that galaxies change from type P to type M with time. We suspect that type P with a plateau might be in an early stage of the bar evolution, while type M with a maximum peak is in a late stage of the bar evolution. Besides, the shape of four peaks at the azimuthal profile $Q_{\rm T}(r_{\rm Qb}, \phi)$ grows from the round shape with a similar width of $\phi$ to a sharp shape with two-to-two pair. It hints that we may estimate the evolutionary stage of bars by investigating the pattern of the ratio map. 
\end{enumerate}

We thank the anonymous referee for careful reading, helpful comments, and insightful guidance, which greatly improved this paper. We also thank Woong-Tae Kim for helpful discussion. We thank the BK21 Plus of the National Research Foundation of Korea (22A20130000179). YHL and MGP acknowledge support by the KASI under the R\&D program supervised by the Ministry of Science, ICT and Future Planning, by the National Research Foundation of Korea to the Center for Galaxy Evolution Research (No.2017R1A5A1070354), and by Basic Science Research Program through the National Research Foundation of Korea (NRF) funded by the Ministry of Education (No.2019R1I1A3A02062242). TK was supported by the Basic Science Research Program through the National Research Foundation of Korean (NRF) funded by the Ministry of Education (No.2019R1A6A3A01092024), and WYS by the grant (2017R1A4A1015178) of the National Research Foundation of Korea.



\end{document}